\documentclass[
  aps,
  prd,
  reprint,
  superscriptaddress,
  nofootinbib,
  longbibliography,
  showkeys,
  floatfix
]{revtex4-2}

\usepackage[T1]{fontenc}

\usepackage{graphicx}
\usepackage{amsmath}
\usepackage{amssymb}
\usepackage{bm}
\usepackage{xcolor}
\usepackage{array}  
\newcolumntype{R}{>{$}r<{$}} 
\newcolumntype{L}{>{$}l<{$}} 

\usepackage{tikz}
\usetikzlibrary{positioning}
\usetikzlibrary{positioning, calc}
\usetikzlibrary{positioning, arrows.meta}
\tikzstyle{box} = [rectangle, draw=black, rounded corners, minimum width=3cm, minimum height=1cm, align=center]
\tikzstyle{arrow} = [thick, -{Latex[length=3mm, width=2mm]}]

\usepackage{silence}
\usepackage{hyperref}
\hypersetup{
  colorlinks=true,
  linkcolor=blue,
  citecolor=blue,
  urlcolor=blue,
  pdftitle={Pad\'e semi-cosmographic reconstruction of residual dark energy in decaying dark matter cosmology: DESI DR2 BAO constraints and mock 21\,cm forecasts },
  pdfauthor={Mohit Yadav, Pankaj Chavan, Tapomoy Guha Sarkar}
}

\newcommand{\LCDM}{\Lambda{\rm CDM}}

\newcommand{\dd}{\mathop{}\!\mathrm{d}}

\begin{document}

\title{Pad\'e semi-cosmographic reconstruction of residual dark energy in decaying dark matter cosmology: DESI DR2 BAO constraints and mock 21\,cm forecasts 
}

\author{Mohit Yadav}
\email{P20210462@pilani.bits-pilani.ac.in}
\affiliation{Department of Physics, Birla Institute of Technology and Science, Pilani, Pilani 333031, Rajasthan, India}

\author{Pankaj Chavan}
\email{P20230079@pilani.bits-pilani.ac.in}
\affiliation{Department of Physics, Birla Institute of Technology and Science, Pilani, Pilani 333031, Rajasthan, India}

\author{Tapomoy Guha Sarkar}
\email{tapomoy1@pilani.bits-pilani.ac.in}
\affiliation{Department of Physics, Birla Institute of Technology and Science, Pilani, Pilani 333031, Rajasthan, India}

\date{\today}

\begin{abstract}
Pure cosmographic reconstructions provide a model-agnostic approach towards constraining cosmic evolution. 
We develop a hybrid phenomenological semi-cosmographic framework that adopts a Pad\'e-rational fraction parametrization of the luminosity distance, and incorporates a phenomenologically motivated two-body decaying dark matter (DDM) sector. In this framework, the effective residual dark-energy equation of state is obtained by subtracting the contributions of DDM, baryons, and radiation from the Pad\'e approximation to the expansion history under the assumptions of flat FLRW cosmology and the standard Friedmann equations. Cosmological observables computed from these semi-cosmographic quantities are fitted to the data to reconstruct cosmological evolution. We present two separate analyses. First, we perform a data fit to the current DESI DR2 BAO measurements. Second, we report a forecast based on an optimistic mock 21\,cm intensity-mapping power spectrum at $z\simeq 1.75$ for a SKA1-MID-like instrument, used jointly with the DESI DR2 BAO data. The forecast serves as proof of concept to demonstrate a possible improvement in model constraints when measurements of the 21\,cm power spectrum complement the BAO background observables.
Posterior constraints on the Pad\'e and DDM parameters are obtained using Markov Chain Monte Carlo (MCMC) analysis, and we present the corresponding equations of state of the massive daughter and the effective reconstructed dark energy. 
\end{abstract}


\maketitle

\section{Introduction}
\label{sec:intro}

Several independent observations, including galactic rotation curves \citep{rubin,sofue,bosma,lelli}, CMBR anisotropies \citep{smoot,dodleson}, large-scale structure surveys \citep{peebles, peacock, weinberg}, supernova observations \citep{riess,perlmutter,betoule,sclonic}, baryon acoustic oscillations \citep{eisenstein,anderson,alam}, weak and strong gravitational lensing \citep{bartlemann,kilbinger,abbott,refsdal,treu}, provide compelling evidence that dark matter (DM) and dark energy (DE) constitute approximately $95\%$ of the total energy budget of the Universe (DM $\sim25\%$ and DE $\sim70\%$). These two components together form a dark sector that plays the dominant role in background cosmological evolution and structure formation. The standard  $\Lambda$CDM cosmological model assumes that dark matter is cold (predominantly responsible for cosmological structure formation) and the cosmological constant ($\Lambda$) is the dark energy candidate responsible for cosmic acceleration. This standard concordance model is able to explain a wide range of cosmological data. However, despite the model's phenomenological success, the $\Lambda$CDM model offers no fundamental account of the dark sector, providing neither an identification of dark matter beyond the Standard Model of particle physics nor a physical understanding of the elusive dark energy component driving cosmic acceleration. Beyond the unresolved theoretical nature of dark matter \citep{davis1985evolution,turner2000dark} and dark energy \citep{scranton2003physical,durrer2008dark,amendola_tsujikawa_2010,Zhao_2017},  persistent observational tensions \citep{Poulin(2019),Niedermann(2019),Sakstein(2020),Ruchika(2020),Banihashemi2(2020),Mena(2020),Alestas(2020)} in cosmological data further suggest that the standard $\Lambda$CDM framework may be incomplete.
In response to these issues confronting the standard $\Lambda$CDM cosmological model, a wide range of theoretical proposals have emerged. These proposals include not only a diverse class of dark energy scenarios \citep{Ratra-Peebles_1988, Steinhardt_1998,PhysRevLett.82.896,scherrer2008thawing} but also alternatives to cold dark matter, such as warm dark matter \citep{bode,viel,dayal}, decaying dark matter \citep{Blackadder2014,blackadder2016cosmological}, and other non-standard dark sector frameworks \citep{amendola2000coupled, Skordis_Coupled_DE_2013}. 
Yet, in the absence of a single framework that consistently reconciles all available observations, there is increasing emphasis on data-driven approaches. The growing number of high-precision cosmological surveys probing the expansion history across a broad redshift range has also made such strategies viable. At the extreme end of model agnosticism lie machine-learning–based methods, such as Gaussian process reconstructions \citep{Holsclaw_2011_GPR,Shafieloo_2012_GPR, Jesus_2024_GPR, Dinda2024_GP_cosmography,Velazquez_mukherjee_GPR_2024, Purba_Mukherjee-Anjan_sen_GPR_2024, Dinda_2025}, which infer cosmological observables directly from data with minimal prior theoretical assumptions.  Such purely data-driven approaches largely exclude physically motivated insights into the underlying dynamical evolution.

{A widely used alternative strategy is cosmography \citep{Weinberg_1972_cosmography}, which shifts the attention from dynamical assumptions to the kinematics of the expanding universe.}

 In this framework, observable quantities such as luminosity distance or the Hubble parameter are expanded as power series in redshift. The expansion coefficients are mapped to kinematic parameters constructed from derivatives of the scale factor  \citep{Visser_2015_cosmography, Dunsby_Luongo_2016_cosmography, Capozziello_2019_cosmography, Busti_2015_cosmography, Visser2005-cosmography, Yang_Aritra_banerjee_2020_cosmography, Aviles_2013_cosmography, Aviles_bravetti_cosmography_2013,Aviles_2012_cosmography_y_variable, Cattoen_2007_convergence, Capozziello_2020_cosmography, pankaj1}. These parameters are then constrained directly using observational data.

The standard cosmographic expansions suffer from a limited radius of convergence, typically breaking down for $z\geq 1$ \citep{Capozziello_2019_cosmography, Cattoen_2007_convergence, Capozziello_2020_cosmography, Lobo_2020, pankaj2,pankaj1}. Adding higher-order terms does not resolve this issue, thereby reducing predictive reliability at high redshifts, precisely where much of the recent supernova and BAO data lie. Re-parameterizations of redshift are sometimes employed to mitigate this limitation \citep{Aviles_2012_cosmography_y_variable, Cattoen_2007_convergence, Capozziello_Ruchika_Anjan_2019_cosmography}.
An improved variant replaces simple Taylor expansions with Padé rational approximants \citep{Pourojaghi2022, Petreca_2024, Wei_2014_cosmography, Capozziello_Ruchika_Anjan_2019_cosmography, Aviles_2014_cosmography, Mehrabi_2018_cosmography, Rezaei_2017_cosmography, Zhou_2016_cosmography, Liu_2021_cosmography, Capozziello_2019_cosmography, Capozziello_2018_cosmography, Benetti_2019_cosmography, Pade_1892}, in which observable quantities are expressed as ratios of two polynomials in $z$.  Such approximations generally possess a larger convergence domain and provide more stable behavior at higher redshifts.

A related pragmatic strategy is \emph{semi-cosmography} \citep{pankaj2,pankaj1}.
In semi-cosmography, one keeps a flexible, data-driven description of the expansion history $H(z)$ reconstructed through cosmography. This description is then embedded within a physically motivated model. For example, one may rely on a data-driven cosmographic approach for the dark energy sector while incorporating existing knowledge of the baryonic matter, dark matter, or radiation component \citep{pankaj2,pankaj1}. In this hybrid phenomenological approach, the effective dark-sector properties are inferred within a framework that keeps explicit assumptions about the background geometry, the expansion-history ansatz, and the matter sector.

The reason for adopting such a framework is that complete agnosticism is not always the most informative choice. Any reconstruction, even a non-parametric one such as Gaussian Process Regression \citep{gp1,gp2,gp3,gp4,gp5}, still contains assumptions through its variables, priors, kernels, smoothness conditions, or background spacetime. The useful question is therefore where the freedom should be placed. Dark matter and dark energy are not equally constrained by present observations. The existence of a clustered matter sector is well established, and many non-standard matter scenarios are already bounded by CMB, lensing, clustering, and large-scale-structure data, whereas the physical origin of dark energy remains less clear. If both sectors are allowed to vary freely, the same departure from $\Lambda$CDM can often be absorbed either into the matter evolution or into an effective residual dark-energy history, making the reconstruction difficult to interpret. Semi-cosmography reduces this ambiguity by keeping the expansion history flexible, but by using physically motivated information in the matter sector to guide the residual dark-energy reconstruction. In the present work, this logic is applied to a two-body DDM scenario. Rather than asking whether a chosen dark energy model fits the observations, we ask what effective residual dark-energy behavior is required once a phenomenologically motivated decaying dark matter sector is included. Thus, the method is best described as a hybrid phenomenological and semi-parametric reconstruction that combines a Pad\'e rational expansion history with a specified two-body DDM sector in a flat FLRW background.

While cold dark matter (CDM) remains the prevailing paradigm, numerous alternative dark matter scenarios have been proposed to address its theoretical and observational limitations \citep{2005PhR...405..279B,2015PNAS..11212249W,2017ARA&A..55..343B,2000PhRvL..84.3760S,2000PhRvL..85.1158H}. Among these, decaying dark matter has drawn particular interest, especially for its potential to ease small-scale structure issues \citep{2001ApJ...546L..77C,2008PhRvD..77f3514B,2010PhRvD..81j3501P}.
There is no compelling a priori reason for dark matter to be perfectly stable, and late-time decay can suppress structure formation, offering a possible joint alleviation of the Hubble and 
$S_8$ tensions \citep{Riess2016,DiValentino2021,Verde2019,Hildebrandt2017,Asgari2021,Abdalla2022,Heymans2021}. Within a semi-cosmographic framework, we consider a decaying dark matter (DDM) model in which a parent dark matter species decays into a massless daughter and a massive daughter. The decay is phenomenologically described by a lifetime and a parameter that controls how the energy is split between the relativistic and massive channels \citep{Blackadder2014,blackadder2016cosmological,abellan2021linear}.
Such decays modify the background evolution and leave imprints on the growth of structure, allowing the parameters to be constrained using both distance measurements and clustering data. Such models have been constrained through multiple observational probes. In this work, we extend earlier treatments to examine in greater detail the cosmological implications of a two-body decaying dark matter scenario with an effective residual dark-energy component within the hybrid phenomenological setup.

In this work, we adopt the following strategy.
We begin by constructing a cosmographic Padé approximant for the luminosity distance $D_L^{\mathcal P}(z)$, from which the  Hubble parameter $H^{\mathcal P} (z)$
is obtained \citep{pankaj2,pankaj1}. Building on this, we develop a semi-cosmographic framework that incorporates the two-body decaying dark matter (DDM) scenario. The energy densities of the parent and daughter components for a spatially flat cosmology are modeled, allowing us to construct an effective residual dark-energy equation of state $w_\phi^{\mathcal P}(z)$, which now depends on both the Pad\'e and DDM parameters. 

The cosmological observables are then obtained using the DDM densities and the equation of state (EoS) of the effective reconstructed dark energy. We then perform two distinct analyses. In the first analysis, we fit the semi-cosmographically computed observables to the current DESI DR2 BAO data \citep{DESI_Colab_2025_DR2}. In the second analysis, we carry out an optimistic proof of concept forecast using a mock 21\,cm intensity-mapping power spectrum at $z\simeq 1.75$ for a SKA1-MID-like interferometer \citep{mohit}\footnote{\url{https://www.skao.int/en}} jointly with the DESI DR2 BAO data. The combined case should be taken as a complementarity study rather than as a prediction for the exact data combination that will be available when such future 21\,cm observations become available. 
This forecast aims to demonstrate a possible tightening of the parameter space when mock 21\,cm intensity mapping complements the BAO background observables. 
The fitted posteriors obtained in the two analyses (DESI DR2 BAO data and the mock forecast with 21\,cm IM) are then used for the residual dark-energy reconstruction and to reconstruct the massive-daughter equation of state. 

The paper is organized as follows. In Section~\ref{subsec:ddm} and\ref{subsec:pade}, we introduce the Pad\'e parametrization and the two-body DDM scenario. Here, we also introduce the residual dark-energy reconstruction within the semi-cosmographic framework.  In section~\ref{sec:obsn_data}, we explicitly discuss the current DESI DR2 BAO data and the mock 21\,cm forecast setup. In Section~\ref{sec:results}, we present, separately, the current constraints from DESI DR2 BAO and the joint forecast when future mock 21\,cm observations are included.  We summarize and conclude our work in Section~\ref{sec:conclusion}.

\section{Formalism: semi-cosmographic framework with two-body DDM and effective residual dark energy}
\label{sec:formalism}

\subsection{Two-body decaying dark matter}
\label{subsec:ddm}
We model the dark matter sector as a minimal extension of standard cold dark matter in which the dominant dark matter species is not perfectly stable \citep{ibarra2013indirect}.
Instead, a non-relativistic ``parent'' particle (label 0) decays into two daughters. One daughter is a massless relativistic particle (label 1), often interpreted as dark radiation, and the other is a massive particle (label 2).
Such a two-body channel is a useful benchmark because it introduces the fewest new parameters while capturing two key physical effects at late times. These are (i) a gradual conversion of matter into radiation-like energy density and (ii) the production of a massive daughter with a non-zero recoil, which can behave as a warm component for some period of time \citep{Blackadder2014,blackadder2016cosmological,vattis2019dark,vattis2019late,abellan2021linear,Vattis}.
Both effects can alter the background expansion and, through free-streaming, can leave characteristic signatures in the growth of structure.

\begin{figure*}
\centering
\includegraphics[width=0.49\textwidth]{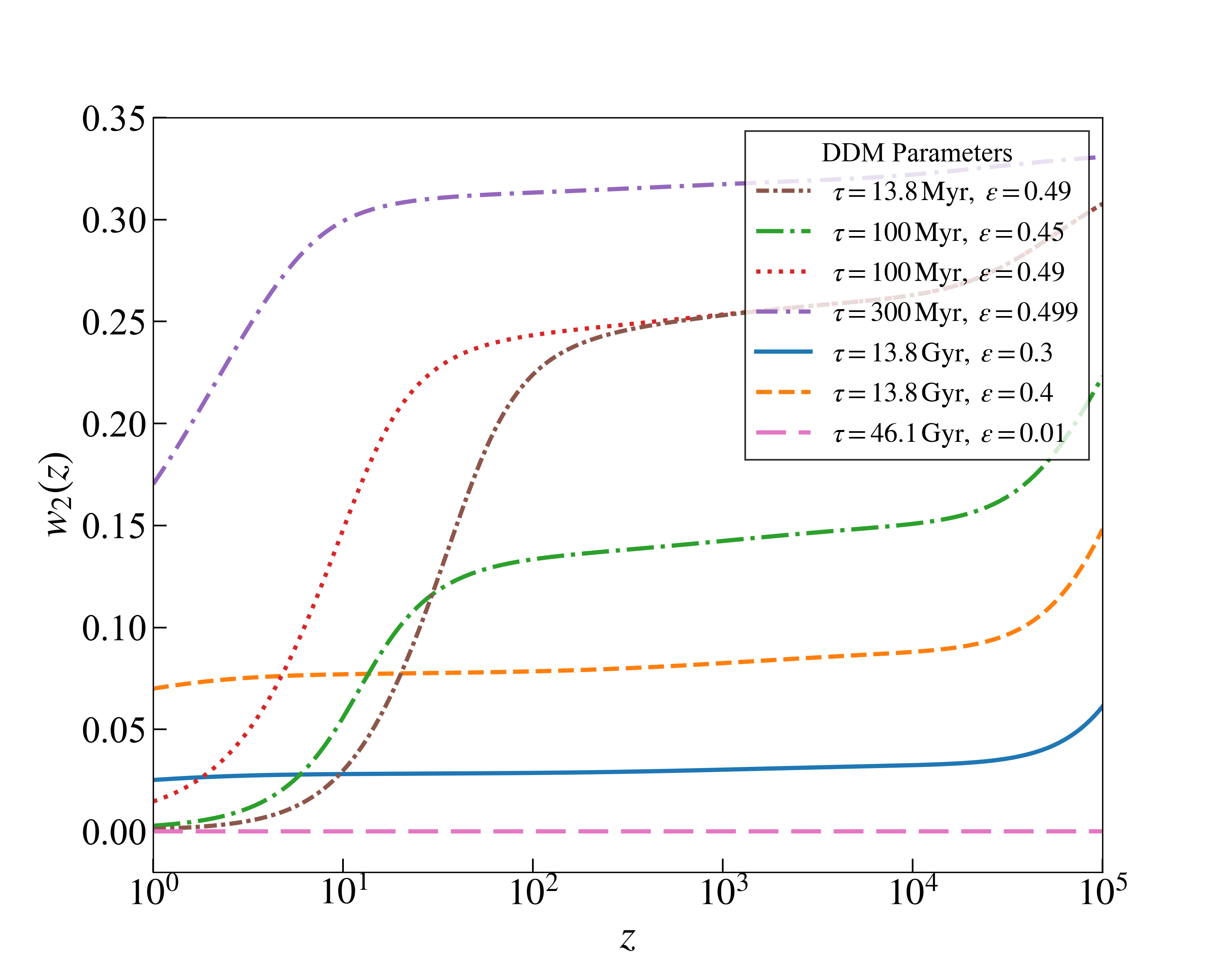}
\includegraphics[width=0.49\textwidth]{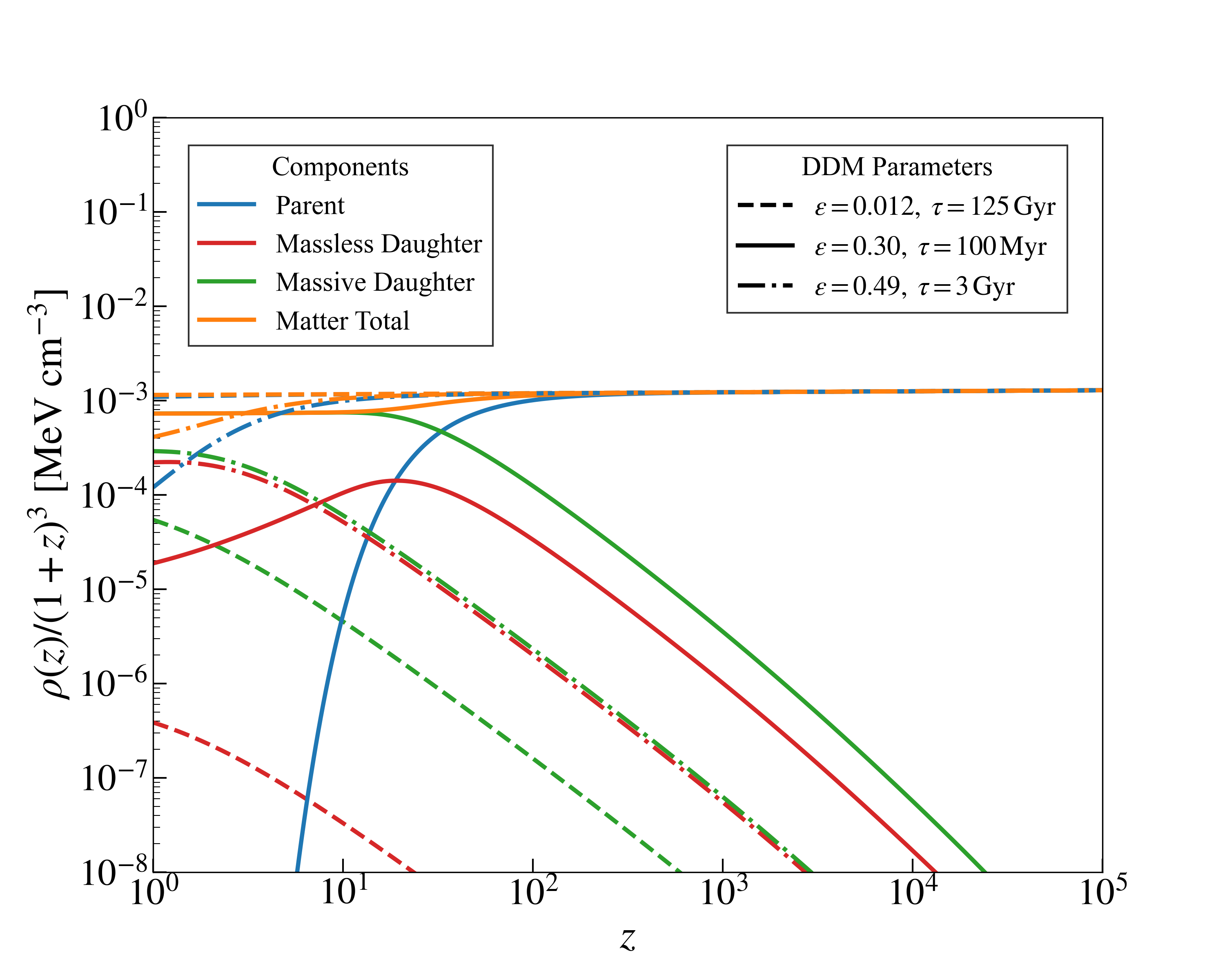}

\caption{{Left:} Massive-daughter EoS $w_2(a)$.
\textit{Right:} Redshift evolution of the rescaled densities $\rho(z)/(1+z)^3$ for the parent dark matter, massless daughter, massive daughter,  and total matter components in the two-body decaying dark matter model.}
\label{fig:w2andden}
\end{figure*}

The decay is characterized by a constant decay rate $\Gamma \equiv 1/\tau$ and a dimensionless parameter $\epsilon$ that controls the partition of the parent's rest-mass energy.
Working in the rest frame of the parent particle, energy--momentum conservation fixes the kinematics of the two daughters.
If $m_0$ and $m_2$ denote the parent and massive-daughter masses, then the fraction of the parent's rest energy carried by the massless daughter can be written as \citep{Blackadder2014,vattis2019dark,abellan2021linear}
\begin{equation}
\epsilon 
= \frac{1}{2}\left(1-\frac{m_2^2}{m_0^2}\right),
\end{equation}
which implies $0 \le \epsilon \le 1/2$ for a physical massive daughter ($m_2 \le m_0$).
The massive daughter is produced with a recoil (or ``kick'') velocity set by the same parameter,
\begin{equation}
\beta = \frac{\epsilon}{1-\epsilon},~~~ \epsilon \simeq \frac{v_{\rm k}}{c},
\end{equation}
so that $\epsilon \ll 1$ corresponds to a small kick. This implies that the massive daughter is nearly cold. A large $\epsilon$, on the other hand, produces a relativistic daughter at birth.
In this way, $\epsilon$ simultaneously controls the amount of dark-radiation injection and the warmness of the massive daughter population.

At the homogeneous (background) level, we treat the parent as a pressureless component, the massless daughter as radiation-like, and the massive daughter as a component with a time-dependent effective equation of state  $w_2(a) \equiv P_2(a)/\rho_2(a)$.
The background energy densities satisfy coupled continuity equations with source terms that encode the conversion of the parent into daughters \citep{Blackadder2014,blackadder2016cosmological,vattis2019dark,abellan2021linear}. These are given by 
\begin{align}
\frac{\mathrm{d} \rho_0}{\mathrm{d} \ln a} &= -3\rho_0 - \frac{\Gamma}{H}\rho_0, \label{eq:ddm0} \\
\frac{\mathrm{d} \rho_1}{\mathrm{d} \ln a} &= -4\rho_1 + \epsilon\,\frac{\Gamma}{H}\rho_0, \label{eq:ddm1} \\
\frac{\mathrm{d} \rho_2}{\mathrm{d} \ln a} &= -3\left(1+w_2\right)\rho_2 + (1-\epsilon)\,\frac{\Gamma}{H}\rho_0,
\label{eq:ddm2}
\end{align}
where $H(a)$ is the Hubble expansion rate.
The first terms on the right-hand side represent the usual redshifting of matter and radiation in an expanding Universe, while the $\Gamma/H$ terms describe decay in cosmic time expressed in the convenient variable $\ln a$.

\begin{figure*}
\centering
\resizebox{\textwidth}{!}{%
\begin{tikzpicture}[
    node distance=1cm,
    box/.style={draw,rounded corners,minimum width=3cm,minimum height=1cm,align=center},
    bigbox/.style={ draw, rounded corners, minimum width=5cm, minimum height=3cm, align=center},
    arrow/.style={->, thick}]

\node (r1) [box, thick] {\textbf{Cosmography}};

\node (r2) [box, below=of r1, thick] {$D_L^{\mathcal{P}}(z) = \frac{\displaystyle c}{\displaystyle H_0}\,\left[\frac{\displaystyle 2\left(\xi^4-a_1\xi^3-(1-a_1)\xi^2\right)}
{\displaystyle b_1\xi^2+c_1\xi + \left(2-a_1-b_1-c_1\right)}\,\right]$ \\ \\ where $\xi=\sqrt{1+z}$ \\ \\  $D_L^{\mathcal P}(z) \equiv D_L^{\mathcal P} \left(z;h,a_1,b_1,c_1 \right)$ with $h=H_0/100$};

\node (r3) [box, below=of r2, thick] {$H^{\mathcal P}(z)= c\left [\frac{d}{dz} \frac{D_L^{\mathcal P}(z)}{(1+z)}\right ]^{-1} $ \\ \\
$H^{\mathcal P}(z) \equiv H^{\mathcal P} \left(z;h,a_1,b_1,c_1 \right)$};

\node (r3pt5) [box, below=of r3, thick] {\textbf{Semi-cosmography}};

\node (r4) [box, below=of r3pt5, thick] {$\displaystyle w_\phi^{\mathcal P}(z) =
\frac{2(1+z)}{3\Omega_\phi^{\mathcal P}(z)}
\frac{d \ln H^{\mathcal P}}{dz}
- \frac{\Omega_r^{\mathcal P}(z) + \Omega_1^{\mathcal P}(z)}
{3\Omega_\phi^{\mathcal P}(z)}
- \frac{w_2(z)\Omega_2^{\mathcal P}(z)}
{\Omega_\phi^{\mathcal P}(z)}
- \frac{1}{\Omega_\phi^{\mathcal P}(z)}$ \\ \\ 
where $\Omega_{\phi}^{\mathcal P}(z)
=
1-\Omega_b^{\mathcal P}(z)-\Omega_r^{\mathcal P}(z)
-\sum_{\alpha=0}^{2}\Omega_\alpha^{\mathcal P}(z)$
\\ \\ $w_\phi^{\mathcal P}(z)\equiv w_{\phi}^{\mathcal P}(z;h,\varepsilon,\tau,\Omega_{dm},a_1,b_1,c_1)$};

\node (r5) [box, below=of r4, thick] {Calculate all observables using the effective reconstructed dark-energy EoS\\
$w_{\phi}^{\mathcal P}(z;h,\varepsilon,\tau,\Omega_{dm},a_1,b_1,c_1)
\rightarrow [D_M/r_d, D_H/r_d, D_V/r_d]$};

\node (r6) [box, below=of r5, thick] {Fit $[D_M/r_d, D_H/r_d, D_V/r_d]$ to data and constrain \\
the DDM and cosmographic parameters $(h,\varepsilon,\tau,\Omega_{dm},a_1,b_1,c_1)$ \\
using Markov Chain Monte Carlo sampling};

\node (leftbox) [bigbox, right=1.5cm of r2, yshift=-0.5cm, thick] {\\
\textbf{DDM scenario with effective residual dark energy} \\ \\
Use instantaneous fractional densities \\ \\
$\displaystyle \Omega_i(z)=\frac{8\pi G\rho_i(z)}{3H^2(z)}$ \\ \\
Spatially flat background \\
$\displaystyle
1=\Omega_b(z)+\Omega_r(z)+\sum_{\alpha=0}^{2}\Omega_\alpha(z)+\Omega_\phi(z)$ \\ \\
Effective residual dark-energy equation of state \\ \\
$\displaystyle w_\phi(z) =
\frac{2(1+z)}{3\Omega_\phi(z)}
\frac{d \ln H}{dz}
- \frac{\Omega_r(z) + \Omega_1(z)}
{3\Omega_\phi(z)}
- \frac{w_2(z)\Omega_2(z)}
{\Omega_\phi(z)}
- \frac{1}{\Omega_\phi(z)}$};

\node (leftbox2) [box, below=of leftbox, thick] {Semi-cosmography $\equiv$ use the same density convention \\
and replace $H(z)$ by $H^{\mathcal P}(z)$ \\ 
Dark matter follows the DDM model with parameters $(h,\varepsilon,\tau,\Omega_{dm})$ \\
Residual dark-energy reconstruction through the effective EoS};

\draw[-{Latex[length=4mm, width=2.5mm]}, very thick] (leftbox) -- (leftbox2);
\draw[-{Latex[length=4mm, width=2.5mm]}, very thick] (leftbox2.south) |- (r4.east);
\draw[-{Latex[length=4mm, width=2.5mm]}, very thick] (r1) -- (r2);
\draw[-{Latex[length=4mm, width=2.5mm]}, very thick] (r2) -- (r3);
\draw[-{Latex[length=4mm, width=2.5mm]}, very thick] (r3) -- (r3pt5);
\draw[-{Latex[length=4mm, width=2.5mm]}, very thick] (r3pt5) -- (r4);
\draw[-{Latex[length=4mm, width=2.5mm]}, very thick] (r4) -- (r5);
\draw[-{Latex[length=4mm, width=2.5mm]}, very thick] (r5) -- (r6);

\end{tikzpicture}%
}
\caption{Schematic flowchart of the hybrid phenomenological semi-cosmographic reconstruction combining a Padé late-time expansion history with a specified two-body DDM sector in flat FLRW geometry.}
\label{fig:Flowchart}
\end{figure*}

The relative weights $\epsilon$ and $(1-\epsilon)$ enforce energy bookkeeping. A fraction $\epsilon$ of the injected energy goes into the massless daughter, and the remainder goes into the massive daughter.
The late-time matter sector relevant for clustering is $\rho_{\rm dm}=\rho_0+\rho_2$, while $\rho_1$ behaves as an additional dark-radiation contribution.

It is useful to note that the parent density admits a simple closed-form solution once the expansion history is specified.
For an initial scale factor $a_{\rm i}$ at which the parent dominates, the dark sector and decay have not yet produced a significant daughter abundance. Therefore, 
\begin{equation}
\rho_0(a)=\rho_0(a_{\rm i})\left(\frac{a_{\rm i}}{a}\right)^3
\exp\!\left[-\Gamma\left(t(a)-t(a_{\rm i})\right)\right],
\end{equation}
where $t(a)$ is the cosmic time implied by the background expansion $H(a)$ \citep{Blackadder2014,blackadder2016cosmological,Vattis}.
This indicates that the parent density is reduced by an exponential decay factor, with the impact controlled by the ratio of the lifetime to the Hubble time.

A distinctive aspect of the two-body scenario is that the massive daughter is not exactly cold.
Each decay injects daughter particles with a fixed physical momentum at the time of production, and the momentum then redshifts as $p \propto a^{-1}$.
Therefore, the daughter population is a superposition of particles produced at different times. Early-produced daughters have had more time to cool and behave nearly as cold matter, while late-produced daughters can remain noticeably warm.
Hence, the massive daughter cannot be described by a constant equation of state.
Instead, one defines an effective $w_2(a)=P_2(a)/\rho_2(a)$ computed from the evolving momentum distribution generated by the decay history as shown below:
\begin{equation}
\label{eq:w2integral}
\begin{aligned}
w_2(a)= {} &\frac{\Gamma\,\beta^2}{3\left(e^{-\Gamma t(a_{\rm i})}-e^{-\Gamma t(a)}\right)} \\
&\times \int_{\ln a_{\rm i}}^{\ln a}\!\dd\ln a'\,
\frac{e^{-\Gamma t(a')}}{H(a')}\\
&\quad\times
\left[\left(\frac{a}{a'}\right)^2(1-\beta^2)+\beta^2\right]^{-1}\,.
\end{aligned}
\end{equation}

In our implementation, $w_2(a)$ is computed self-consistently from the decay-time distribution, leading to the integral expression in Eq.~\eqref{eq:w2integral}. Once $(\epsilon,\tau)$ and the expansion history are fixed, $w_2(a)$ is determined.
The parameters $(\epsilon,\tau)$ control two qualitatively different limits.
If $\Gamma t_0 \ll 1$ (very long lifetime) or $\epsilon \to 0$ (negligible kick), then $\rho_0$ redshifts almost as $a^{-3}$ and the daughter remains effectively cold ($w_2 \approx 0$), so the scenario reduces smoothly to the standard CDM case.
If the lifetime is comparable to the age of the Universe and $\epsilon$ is not extremely small, then a non-negligible fraction of matter is converted into dark radiation and into a warm massive daughter.

This reduces the late-time matter abundance relative to stable CDM and can suppress the growth of structure below a free-streaming scale set by the kick velocity and the decay epoch.
The left panel of Fig.~\ref{fig:w2andden} shows the variation of the equation of state for the massive daughter for a host of DDM model parameters. When $\epsilon$ is small and $\tau H_0\gg1$, we find that  $w_2 \sim 0$ and thus the massive daughter behaves like cold dark matter. However, for large $\epsilon$, $w(z)$  increases with redshift. This rise is steep at low redshifts and more gradual at higher redshifts. 
The right panel of Fig.~\ref{fig:w2andden} shows the variation of the parent dark matter and daughter densities obtained by numerically integrating equations ~\eqref{eq:ddm0},~\eqref{eq:ddm1}, and~\eqref{eq:ddm2}.
We have considered three DDM models with parameters $(\epsilon, \tau) =(0.012,125 $~Gyr$)$, $(0.30,100 $~Myr$ )$ and $(0.49,3 $~Gyr$)$,
 respectively.  The figure shows how the total matter is distributed among the three populations. The first model, in which the decay time is very large and $\epsilon$ is small, closely mimics cold dark matter. The other two models represent substantial departures from the CDM scenario. 

\subsection{Semi-cosmography in the DDM scenario}
\label{subsec:pade}

In a standard Pad\'e cosmographic approach, the background cosmological evolution is described through the kinematics of the Universe. Observable quantities such as the Hubble expansion rate $H(z)$, the angular diameter distance $D_A(z)$, and the luminosity distance $D_L(z)$ are expanded through Pad\'e rational approximants \citep{Pade_1892, Capozziello_2019_cosmography, Capozziello_2018_cosmography, Benetti_2019_cosmography}. These Pad\'e-approximated observables are then fitted directly to the observational data to constrain kinematical parameters like the Hubble constant $H_0$, the deceleration parameter $q_0$, and the jerk parameter $j_0$. This traditional cosmographic approach is purely kinematical in nature and does not provide a direct natural way to include parameters related to the dynamics of the Universe.

In our study, we adopt a Pad\'e-type rational approximation for the luminosity distance in terms of the variable $\xi \equiv \sqrt{1+z}\,$ \citep{Saini_2000}, defined as
\begin{equation}
D_L^{\mathcal{P}}(z) = \frac{c}{H_0}\,\left[\frac{2\left(\xi^4-a_1\xi^3-(1-a_1)\xi^2\right)}
{b_1\xi^2+c_1\xi + \left(2-a_1-b_1-c_1\right)}\,\right].
\label{eq:DL_pade_def}
\end{equation}
This choice for the luminosity distance is not arbitrary. In the high- redshift limit ($z\gg1$), the Hubble expansion rate $H^{\mathcal{P}}(z)$ obtained from $D_L^{\mathcal{P}}(z)$ through
\begin{equation}
H^{\mathcal{P}}(z)=c\left[\frac{\dd}{\dd z}\left(\frac{D_L^{\mathcal{P}}(z)}{1+z}\right)\right]^{-1}
\label{eq:Hubble_luminosityDist_relation}
\end{equation}
reproduces an expansion history similar to a matter- dominated Universe. In the low-redshift limit ($z\rightarrow0$), $H^{\mathcal{P}}(z)$ also has the desired asymptotic behavior \citep{Saini_2000}. Instead of relating the Pad\'e parameters $a_1,b_1,c_1$ to the usual kinematical parameters, we treat them as free model parameters to be constrained using the observed data.

The derived Pad\'e-approximated expansion rate $H^{\mathcal{P}}(z)$ is a kinematically determined function. We embed it in a dynamical framework within the decaying dark matter scenario. To keep the reconstruction consistent, we use a single density convention throughout the equation of state derivation. All density variables that appear below are defined as instantaneous fractional densities
\begin{equation}
\Omega_i(z)
=
\frac{8\pi G\rho_i(z)}{3H^2(z)},
\label{eq:omega_instant_definition}
\end{equation}
where $i$ denotes baryons, radiation, the parent dark matter particle, the massless daughter, the massive daughter, and the effective residual dark-energy component. With this convention, the spatially flat background satisfies
\begin{equation}
1
=
\Omega_b(z)+\Omega_r(z)
+\sum_{\alpha=0}^{2}\Omega_\alpha(z)
+\Omega_\phi(z).
\label{eq:flatness_instant}
\end{equation}

The parent particle and baryons are pressureless ($w_0 = w_b =0$). Radiation and the massless daughter both have the equation of state $w_r = w_1= 1/3$. The massive daughter contributes through its effective equation of state $w_2(z)$, while the effective residual dark-energy component contributes through its reconstructed equation of state $w_\phi(z)$. The continuity equations \eqref{eq:ddm0}, \eqref{eq:ddm1} and \eqref{eq:ddm2}, together with the Friedmann equation for a flat Universe, yield
\begin{equation}
\begin{aligned}
\frac{2(1+z)}{3}
\frac{d\ln H}{dz}
={}&
1
+
\frac{\Omega_r(z)+\Omega_1(z)}{3}
\\
&+
w_2(z)\Omega_2(z)
+
w_\phi(z)\Omega_\phi(z).
\end{aligned}
\label{eq:h_derivative_ddm}
\end{equation}
The source terms in the DDM continuity equations cancel when the total energy balance is written, so only the total pressure enters this relation. Solving Eq.~\eqref{eq:h_derivative_ddm} for $w_\phi(z)$ gives
\begin{equation}
\begin{aligned}
w_\phi(z) ={}&
\frac{2(1+z)}{3\Omega_\phi(z)}
\frac{d \ln H}{dz}
- \frac{\Omega_r(z) + \Omega_1(z)}
{3\Omega_\phi(z)}
\\
&- \frac{w_2(z)\,\Omega_2(z)}
{\Omega_\phi(z)}
- \frac{1}{\Omega_\phi(z)}.
\end{aligned}
\label{eq:wphi_ddm_clean}
\end{equation}
For the semi-cosmographic analysis,  $H(z)$ is replaced by $H^{\mathcal P}(z)$ and $\Omega_i(z)$ by $\Omega_i^{\mathcal P}(z)$, where
\begin{equation}
\begin{aligned}
\Omega_b^{\mathcal P}(z)
&=\frac{\Omega_{b0}(1+z)^3}
{\left[E^{\mathcal P}(z)\right]^2},
~~
\Omega_r^{\mathcal P}(z)
= \frac{\Omega_{r0}(1+z)^4}
{\left[E^{\mathcal P}(z)\right]^2}, ~{\rm and }
\\
\Omega_\alpha^{\mathcal P}(z)
&=
\frac{8\pi G\rho_\alpha(z)}
{3H_0^2\left[E^{\mathcal P}(z)\right]^2}.
\end{aligned}
\label{eq:omega_pade_components}
\end{equation}
The effective residual dark-energy density is then defined by
\begin{equation}
\Omega_{\phi}^{\mathcal P}(z)
=
1
-
\Omega_b^{\mathcal P}(z)
-
\Omega_r^{\mathcal P}(z)
-
\sum_{\alpha=0}^{2}\Omega_\alpha^{\mathcal P}(z).
\label{eq:omega_phi_pade}
\end{equation}
Thus, the equation of state of the effective reconstructed dark energy is given by 
\begin{equation}
\begin{aligned}
w_\phi^{\mathcal P}(z) ={}&
\frac{2(1+z)}
{3\Omega_\phi^{\mathcal P}(z)}
\frac{d \ln H^{\mathcal P}}{dz}
-
\frac{\Omega_r^{\mathcal P}(z)+\Omega_1^{\mathcal P}(z)}
{3\Omega_\phi^{\mathcal P}(z)}
\\
&-
\frac{w_2(z)\,\Omega_2^{\mathcal P}(z)}
{\Omega_\phi^{\mathcal P}(z)}
-
\frac{1}
{\Omega_\phi^{\mathcal P}(z)}.
\end{aligned}
\label{eq:wphi_pade_clean}
\end{equation}

The replacement of $H(z)$ by  $H^{\mathcal P}(z)$ and $\Omega_i(z)$ by $\Omega_i^{\mathcal P}(z)$ implies that $w_\phi^{\mathcal P}(z)$ is obtained from a residual dark-energy reconstruction and is not a dynamical prediction of the DDM model itself. It is conditional on the Pad´e expansion history, the assumed DDM sector, flat FLRW geometry, and the standard Friedmann equation.
The equation of state of the effective reconstructed dark energy, $w_\phi^{\mathcal P}(z)$, is determined jointly by the DDM parameters $(h,\varepsilon,\tau,\Omega_{\rm dm})$ through the DDM densities $\rho_\alpha(z)$ and $w_2(z)$, and by the cosmographic parameters $(a_1,b_1,c_1)$ through $E^{\mathcal P}(z)$. The parameters $(\varepsilon,\tau)$ govern deviations from the cold dark matter limit, while $(a_1,b_1,c_1)$ control the reconstructed expansion history.

In the semi-cosmographic approach, all cosmological observables are obtained using the effective reconstructed dark-energy equation of state $w_\phi^{\mathcal P}(z)$. These observables are then fitted to the data to constrain the DDM parameters and the Pad\'e parameters. Fig.~\ref{fig:Flowchart} shows the schematic flowchart of the semi-cosmographic method.

\begin{figure*}
\centering
\includegraphics[width=0.49\textwidth]{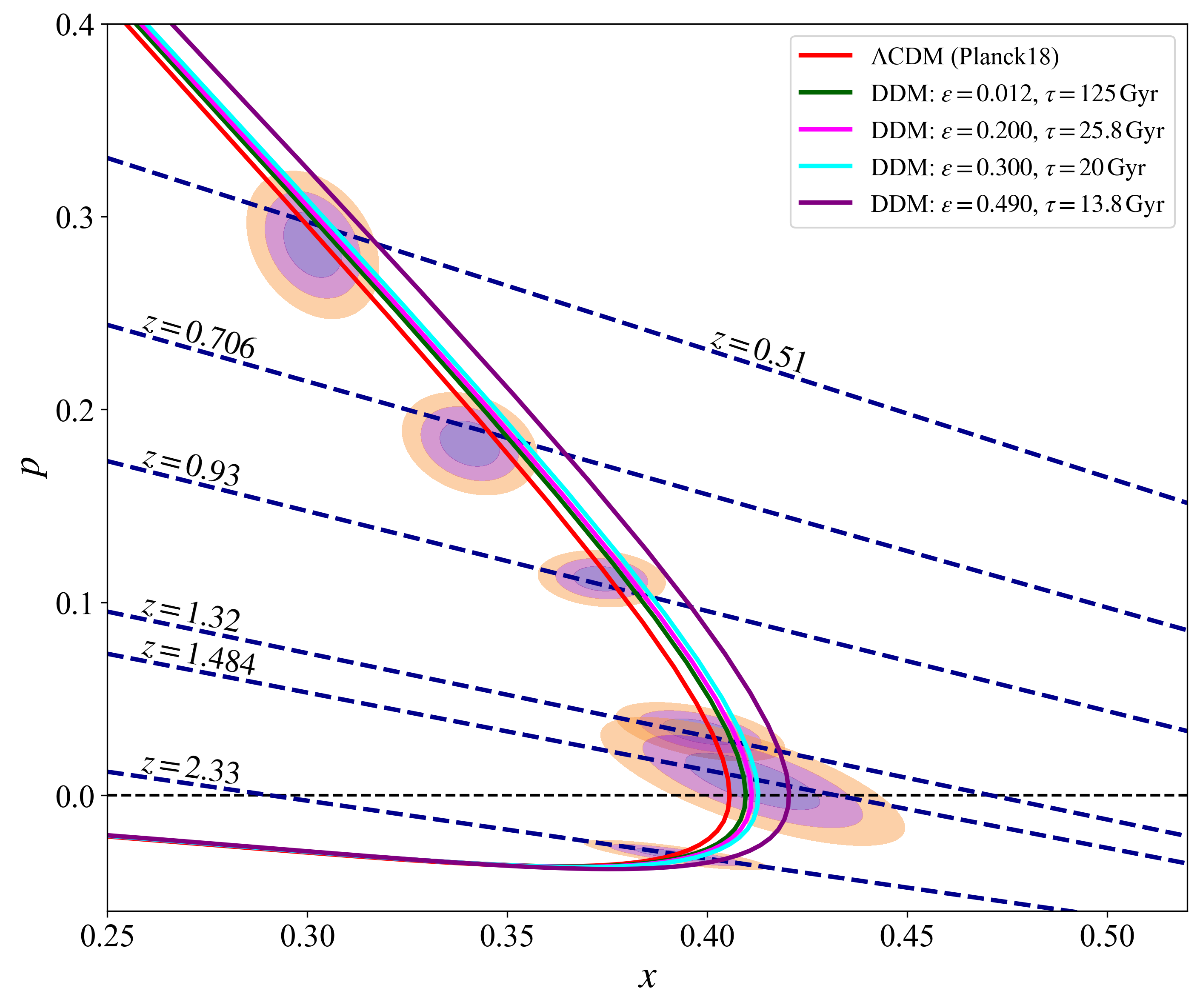}\hfill
\includegraphics[width=0.49\textwidth]{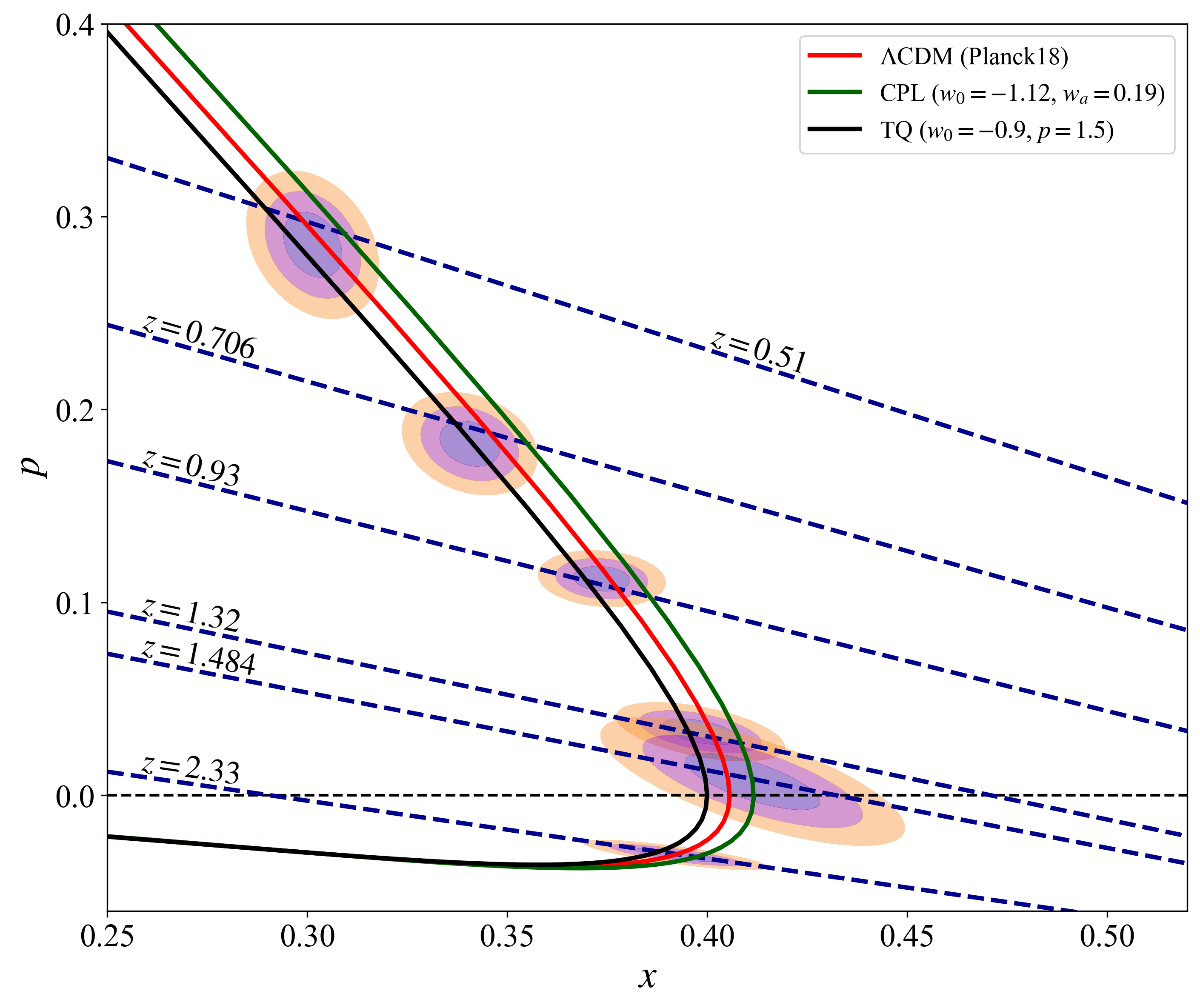}
\caption{The cosmological evolution in $(x,p)$ phase-space.
\textit{Left:}  Here we assume the dark energy to be a cosmological constant and study the phase trajectories for different decay-parameter combinations (coloured curves) $(\epsilon,\tau)$ for a two-body DDM scenario.
\textit{Right:} The dark matter is considered cold, and trajectories are shown for different dark energy models: Chevallier--Polarski--Linder (CPL) and thawing quintessence (TQ).
In both panels, the red-coloured trajectory represents the Planck18 $\Lambda$CDM model. The coloured ellipses show the DESI DR2 BAO projections mapped onto the $(x,p)$ plane (68\%, 95\%, and 99.7\% credible regions).
The blue dashed lines are constant-redshift consistency lines implied by spatial flatness,
$x+(1+z)p = 1/E(z)$, shown at $z=0.51,\,0.706,\,0.93,\,1.32,\,1.484$ and $2.33$.
The intersection of the horizontal dashed line $p=0$ with the trajectories corresponds to the maxima of $D_A(z)$.
}

\label{fig:xp_side_by_side}
\end{figure*}

\section{Current data and optimistic forecast setup} 
\label{sec:obsn_data}
To estimate the DDM and Pad\'e parameters, reconstruct the dark matter evolution, and carry out the residual dark-energy reconstruction, we have used BAO data from DESI DR2 and a futuristic mock 21\,cm intensity mapping mock data with a SKA1-MID-like radio telescope. 

\begin{table*}
\centering
\begin{tabular}{lccccc}
\hline
\hline
Tracer & $z_{\rm eff}$ & $D_M/r_d$ & $D_H/r_d$ & $r_{M,H}$ & $D_V/r_d$ \\
\hline

BGS & 0.295 & --- & --- & --- & $7.942 \pm 0.075$ \\

\hline

LRG1 & 0.510 
& $13.588 \pm 0.167$ 
& $21.863 \pm 0.425$ 
& $-0.459$ 
& --- \\

\hline

LRG2 & 0.706 
& $17.351 \pm 0.177$ 
& $19.455 \pm 0.330$ 
& $-0.404$ 
& --- \\

\hline

LRG3+ELG1 & 0.934 
& $21.576 \pm 0.152$ 
& $17.641 \pm 0.193$ 
& $-0.416$ 
& --- \\

\hline

ELG2 & 1.321 
& $27.601 \pm 0.318$ 
& $14.176 \pm 0.221$ 
& $-0.434$ 
& --- \\

\hline

QSO & 1.484 
& $30.512 \pm 0.760$ 
& $12.817 \pm 0.516$ 
& $-0.500$ 
& --- \\

\hline

Ly$\alpha$ & 2.330 
& $38.988 \pm 0.531$ 
& $8.632 \pm 0.101$ 
& $-0.431$ 
& ---\\

\hline
\hline
\end{tabular}

\caption{The mean and standard deviation data for DESI DR2 BAO observables used in our analysis \citep{DESI_Colab_2025_DR2}. We have adopted the full $13\times13$ covariance matrix from \url{https://github.com/CobayaSampler/bao_data/tree/master/desi_bao_dr2}.}
\label{tab:DESIdata}
\end{table*}

\subsection{BAO data}

We adopt the baryon acoustic oscillation (BAO) data from DESI DR2 \citep{DESI_Colab_2025_DR2} in our analysis. The survey provides measurements of the observables $\tilde{D}_M=D_M/r_d, \tilde{D}_H=D_H/r_d, \tilde{D}_V=D_V/r_d$ and the ratio $\tilde{D}_M/\tilde{D}_H$ for various large-scale tracers. 
The quantities $\tilde{D}_M$ and $\tilde{D}_H$ are related to the transverse and radial comoving distances, respectively. 
The quantity $\tilde D_V$, called the volume-averaged distance, provides a combined isotropic measure of radial and transverse information, while the anisotropic information is contained within the ratio $\tilde{D}_M/\tilde{D}_H$ through the Alcock-Paczynski effect.

These BAO observables are defined in terms of the Hubble expansion rate $H(z)$ and the angular diameter distance $D_A(z)$ as
\begin{equation}
\tilde{D}_M(z) = \frac{c}{r_d} \int_0^z \frac{dz'}{H(z')},~~~\tilde{D}_H(z) = \frac{c}{H(z) r_d},
\end{equation}
and
\begin{equation}
\tilde{D}_V(z) = \frac{c}{r_d} 
\left[ (1+z)^2 D_A^2(z)\frac{cz}{H(z)} \right]^{1/3}.
\end{equation}
Here $c$ is the speed of light in vacuum and $r_d$ is the sound horizon distance at the drag epoch $z_d\approx1060$. 

We consider an independent isotropic distance measure ($\tilde{D}_V$) for the Bright Galaxy Survey (BGS) bin at $z_{\rm eff} = 0.295$. We also consider other non-BGS tracers like Luminous Red Galaxies (LRG), Emission Line Galaxies (ELG), Quasi-Stellar Objects (QSO), such as quasars and Lyman alpha forest (Ly-$\alpha$) in six additional redshift bins $z_{eff} = 0.51, 0.706, 0.934, 1.321, 1.484$ and $2.330$. For these non-BGS tracers, we consider the anisotropic distance measures ($\tilde{D}_M, \tilde{D}_H$). The complete data set, therefore, comprises a total of 13 measurements along with a $13\times13$ covariance matrix. The data set used in this work is summarized in Table \ref{tab:DESIdata}.

In our analysis, we use $r_d=147.21\pm0.23$ from CMB constraints \citep{Planck2018}. Given the sound horizon distance $r_d$, the BAO observables $\tilde{D}_M, \tilde{D}_H, \tilde{D}_V$ and the ratio $\tilde{D}_M/\tilde{D}_H$ can provide independent measurements of the angular diameter distance $D_A(z)$ and the Hubble expansion rate $H(z)$. However,  these quantities are not independent. For a flat, homogeneous, and isotropic Universe, they are related through a consistency relation
\begin{equation}
D_A(z) = \frac{c}{1+z} \int_0^z \frac{dz'}{H(z')}.
\label{Eq: Da_H_relation}
\end{equation}
We define dimensionless  phase-space variables
\begin{equation}
x(z) = \frac{H_0 D_A(z)}{c}, \qquad p(z) = \frac{dx}{dz}.
\end{equation}
In terms of phase-space variables $x$ and $p$, the consistency condition is given by
\begin{equation}
x + (1+z)p = \frac{1}{E(z)},
\label{Eq: consistence_xp}
\end{equation}
where $E(z) = H(z)/H_0$.
This relation represents a straight line in $(x,p)$ phase-space for a given redshift $z$.

The cosmology-independent slope of this straight line is $-(1+z)^{-1}$, and the intercept on the $p$-axis depends on the expansion history through $E(z)$.
 In the $(x,p)$ phase-space, the background cosmological evolution is a unique trajectory parametrized by redshift. All trajectories originate at $(x,p) = (0,1)$ at the present epoch ($z=0$) and asymptotically approach $(0,0)$ (Big Bang) as $z \to \infty$. The physically allowed values of $(x,p)$ at a given redshift, therefore, must lie on the corresponding consistency line. This appears as a point of intersection of a phase trajectory and a consistency line at a given redshift.

Fig.~\ref{fig:xp_side_by_side} shows the background cosmological evolution in the $(x,p)$ phase-space. The left panel isolates the effect of variations in the decaying dark matter model parameters while assuming the cosmological constant ($\Lambda$) to be the dark energy candidate. The right panel shows different dark energy models, such as Chevallier--Polarski--Linder (CPL) \citep{CHEVALLIER_2001}, Thawing quintessence (TQ) \citep{scherrer2008thawing}, considering the dark matter to be cold. 
In these figures, we have superimposed the actual BAO measurements from DESI DR2 \citep{DESI_Colab_2025_DR2} at six redshifts.

The BAO measurements, along with their covariance, are transformed into phase-space variables $(x,p)$. At each redshift, the uncertainties in the observations appear as an ellipse in the $(x,p)$ plane. These error ellipses are shown in both panels. At low redshifts, the ellipses appear tilted, while at higher redshifts they tend to align closely with the consistency lines. The redshift dependence of this behavior can be understood by considering the variation in the consistency relation
\begin{equation}
\delta x + (1+z)\,\delta p = -\frac{\delta E}{E^2}.
\end{equation}
At low redshifts, $E(z) \sim \mathcal{O}(1)$, which indicates that  $\delta E/E^2$ has a significant contribution, allowing larger fluctuations in the direction transverse to the direction of consistency (Eq.~(\ref{Eq: consistence_xp})). This results in the tilt of the error ellipses with respect to the consistency line.
At high redshifts,  $E(z) \gg 1$, whereby  $\delta E/E^2 \ll 1$. In this limit, the ellipses are aligned along the consistency direction.
These ellipses provide a common reference against which various cosmological models can be assessed. The extent to which a given trajectory intersects or deviates from the observational ellipses at fixed redshift provides a stringent diagnostic for model assessment.

For a given redshift, the consistency relation defines a straight line in the $(x,p)$ with a model-independent slope, but with an intercept that depends explicitly on the expansion history through $E(z)$.
This implies the existence of different consistency lines corresponding to different cosmological models. These lines have the same slope, but are shifted along the vertical $p$-axis. The intersection of the phase trajectory with the consistency line for a given model determines the predicted $(x,p)$ at that redshift, which can then be directly compared with the observational ellipses to eliminate a model.

In this work, the residual dark-energy reconstruction is carried out within a hybrid phenomenological semi-cosmographic framework that explicitly assumes a two-body DDM sector and a Pad\'e expansion history. We use a semi-cosmographic data-driven approach to constrain cosmic evolution as described in the last section.

\subsection{Mock 21\,cm intensity-mapping forecast}

In the post-reionization Universe, the diffuse intergalactic medium is mostly ionized, and the observed 21\,cm emission is expected to arise mainly from dense self-shielded regions associated with damped Lyman-$\alpha$ absorbers (DLAs). These systems are believed to contain the bulk of the neutral hydrogen at $z<6$, with column densities above $2\times10^{20}\,{\rm cm}^{-2}$, and are therefore the dominant sources of the post-reionization HI signal \citep{xhibar,xhibar1,xhibar2} seen in emission. Since intensity mapping probes the collective large-scale emission rather than individual emitters, the discrete nature of the DLA population is usually ignored at leading order. This is a reasonable approximation when the effective number density of HI-bearing systems is large, although it also means that the resulting forecast does not include an explicit shot-noise contribution.

At these redshifts, the spin temperature of neutral hydrogen is expected to be much larger than the CMB temperature, so the 21\,cm line is seen in emission against the background radiation field. Observations of Lyman-$\alpha$ absorption systems also indicate that the mean neutral fraction remains roughly constant in the post-reionization era, with a fiducial value $\bar{x}_{\rm HI}\simeq 2.45\times10^{-2}$ that is commonly adopted in large-scale 21\,cm forecasts \citep{proch05}. The HI distribution is expected to trace the underlying dark matter field with a bias $b_T(k,z)$, which is approximately linear and scale-independent on large scales, but becomes scale-dependent on smaller scales \citep{bagla20,Sarkar_2016,Guha_Sarkar_2012}. Numerical studies show that the large-scale HI bias increases with redshift, while the small-scale bias rises more steeply because HI is preferentially hosted by more massive haloes \citep{Mar_n_2010}. 

In the present work, we adopt a simulation-based fitting form for $b_T(k,z)$, as described in \citep{Sarkar_2016}, and use it as a fiducial input for the mock 21\,cm analysis. This fitting form was calibrated in a standard CDM setting, and in the present DDM analysis, it should therefore be understood as an external astrophysical approximation rather than a fully self-consistent prediction. The approximation is mainly justified because, on the large linear scales relevant to most of the cosmological information, the HI field is still expected to behave as a smoothly biased tracer of the matter distribution. 
Since the post-reionization HI bias is obtained as an HI-mass-weighted average of the halo bias, any change in the halo mass function, halo bias, or HI-halo mass relation (how HI is distributed in the halos)  will propagate into the 21\,cm power spectrum.
The DDM affects the 21\,cm power spectrum through (1) the modified matter power spectrum, (2) the modified growth rate $f(z)$, and (3)  the modified HI bias. In our current treatment, the first two effects are included, but the possible DDM-induced modification of the HI bias is not modeled independently.

Physically, DDM suppresses structure formation by reducing the dark matter density and weakening late-time growth. This changes the halo mass function, especially at the high-mass end. If the HI prescription assigns an HI mass approximately proportional to halo mass, then the reduction of massive, highly biased halos would tend to lower the HI-weighted bias and further suppress the 21\,cm power spectrum. On the other hand, for more realistic HI prescriptions, where most HI resides in intermediate-mass halos rather than in clusters, the change in $b_T$ may be smaller and could even be partially compensated by changes in the relative abundance of HI-hosting halos.

Thus, the use of a CDM-calibrated HI bias should be regarded as a baseline approximation. A fully self-consistent treatment would require recalculating the halo mass function, halo bias, and HI-halo relation (the prescription for assigning HI to halos)  for the DDM cosmology, preferably using DDM-specific simulations or a modified halo-model prescription. We note that our forecasts, as conditional on the adopted CDM-calibrated HI bias, may be affected by this modeling uncertainty.

One may adopt an alternative approach by treating the bias as a set of external astrophysical nuisance parameters and marginalizing over them in the MCMC analysis. For the adopted scale- and redshift-dependent fiducial bias model, the nuisance parameters would be the coefficients of the polynomial fit $b_T(k,z)$ obtained from CDM-based simulations \citep{Sarkar_2016}. These expansion coefficients may be allowed to vary and treated as parameters along with the DDM parameters. An earlier work using a simpler Fisher-matrix approach has, however, shown that when the bias parameters are marginalized, the constraints on $(\Gamma, \epsilon)$ are severely degraded \cite{mohit}, rendering the forecast unable to make any statement about DDM cosmology.
For this reason, we have not treated the CDM-calibrated bias parameters as free nuisance parameters to be marginalized over. We thus acknowledge the approximate nature of our forecasts owing to this incomplete modeling.

We also note that several other observational issues may, in fact, degrade the constraints significantly. 
We use an optimistic baseline forecast of 21\,cm intensity-mapping (IM) power spectrum as a complement to the geometric information from BAO, because it probes the growth of structure.
The redshifted 21\,cm signal from the post-reionization epoch is a powerful cosmological probe \citep{poreion0,poreion1,poreion2,poreion3,poreion4,poreion5,poreion6,poreion7,poreion8,poreion9,poreion10,poreion11,poreion12} and several radio telescopes aim to measure this signal \citep{mohit}. 
In the post-reionisation regime ($z\sim 1$--$3$), the 21\,cm signal traces the large-scale distribution of dark matter \citep{poreion0,param1, Bull_2015, param2, param3, param4}. 
Thus, the post-reionization 21\,cm power spectrum is a probe of dark matter clustering. 

Since precision measurements of the 21\,cm power spectrum at the redshifts of interest are not yet available, we construct an optimistic mock anisotropic 21\,cm power-spectrum data set and adopt noise estimates for a future observation with an SKA1-MID-like radio interferometer. We use this as an additional term in the joint likelihood in the joint analysis of parameters.

In linear theory, the redshift-space 21\,cm power spectrum at redshift $z$ is modeled as \citep{param3,bharad04}
\begin{equation}
\begin{aligned}
P_{HI}(k,\mu,z)={}&
\frac{C^2(z)}{\alpha_\perp^2(z)\,\alpha_\parallel(z)}\,
\,\Bigl[1+\beta_T(z)\,\tilde{\mu}^2\Bigr]^2 \\
&\times
P(\tilde{k},z)\,G_{\rm FoG}(\tilde{k},\tilde{\mu},z).
\end{aligned}
\label{eq:P21_model}
\end{equation}
We have adopted the  Alcock-Paczy\'nski (AP) \citep{AP1979,lopez2014alcock} rescaling between the fiducial and trial cosmologies  by defining
\begin{equation}
\begin{aligned}
\alpha_\parallel(z)&=\frac{H^{\rm f}(z)}{H(z)}, &
\alpha_\perp(z)&=\frac{D_A(z)}{D_A^{\rm f}(z)},\\
F(z)&\equiv \frac{\alpha_\parallel(z)}{\alpha_\perp(z)},
\end{aligned}
\end{equation}
Here, $H^f$ and $D_A^f$ are the Hubble parameter and angular-diameter distance, respectively, in a fiducial cosmology. The AP-remapped variables $(\tilde{k},\tilde{\mu})$ are given by 
\begin{equation}
\begin{aligned}
\tilde{k}&=\frac{k}{\alpha_\perp(z)}
\sqrt{1+\mu^2\left[F^{-2}(z)-1\right]},
\\
\tilde{\mu}^2&=\frac{\mu^2}{F^2(z)+\mu^2\left[1-F^2(z)\right]}.
\end{aligned}
\label{eq:AP_kmu}
\end{equation}
The redshift-space distortion parameter $\beta_T$ is given by 
\begin{equation}
\begin{aligned}
\beta_T(k,z)&=\frac{f_g(k,z)}{b_T(k,z)},\\
f_g(z)&\equiv \frac{\dd\ln D_+}{\dd\ln a}.
\end{aligned}
\end{equation}
Here, $D_+(z)$ denotes the linear growing mode of density perturbations and $b_T(k,z)$ is the effective HI bias.
The overall brightness-temperature normalization is given by  \citep{bharad04,param3,masui2013measurement}
\begin{equation}
\begin{aligned}
C(z)={}& 4.0\,{\rm mK}\;\bar{x}_{\rm HI}\,b_T(k,z)\,(1+z)^2\\
&\times
\left(\frac{\Omega_{\rm b0}h^2}{0.02}\right)
\left(\frac{0.7}{h}\right)
\left(\frac{H_0}{H(z)}\right).
\end{aligned}
\label{eq:CT_def}
\end{equation}
We fix the mean neutral fraction $\bar{x}_{\rm HI}$ to a standard post-reionization fiducial value of $2.45\times10^{-2}$ in our forecasts \citep{proch05}.
To account for small-scale velocity damping along the line of sight, we include a Finger-of-God suppression factor \citep{jackson1972critique}
\begin{equation}
G_{\rm FoG}(k,\mu,z)=\left[1+\frac{k^2\mu^2\sigma_p^2}{2}\right]^{-2},
\label{eq:fog}
\end{equation}
where $\sigma_p$ is the velocity-dispersion parameter that we keep fixed in the mock analysis.

For the power spectrum $P(k, z)$, we use the public \texttt{DMemu} package and specifically its two-body decaying dark matter module \texttt{TBDemu} \url{https://github.com/jbucko/DMemu}.
The emulator is based on gravity-only $N$-body simulations of two-body decaying dark matter. It predicts the nonlinear suppression response \citep{Potter,Peters}
\[
S_{\rm DDM}(k,z,\epsilon,\tau)
=
P(k,z)/P_{\Lambda{\rm CDM}}(k,z),
\]
which is then applied to the baseline nonlinear $\Lambda$CDM matter power spectrum.
The emulator accounts for suppression due to decay kinematics and free-streaming of the daughter particles. A fully self-consistent perturbation calculation would, however, require a modified Einstein-Boltzmann treatment, which is beyond the scope of the present work. We state that our subsequent emulator-based forecast-level treatment is not a fully self-consistent calculation, including DDM perturbations, Baryonic effects, and a DDM-dependent HI bias model. We note that the posterior region used in the subsequent forecasts lies within the calibrated parameter and redshift range of the emulator. We also note that parameter proposals outside the calibrated emulator domain are assigned zero posterior weight and therefore rejected during MCMC sampling. The calibrated scale range of the emulator is $k<6\,h\,{\rm Mpc}^{-1}$ \citep{Peters}.

We model instrumental uncertainties with the standard thermal-noise power spectrum for an interferometric array \citep{mcquinn2006cosmological}.
For an observed wavelength $\lambda=0.21(1+z)\,{\rm m}$ and $\nu_{21}=1420.4~{\rm MHz}$, we use
\begin{equation}
\begin{aligned}
N_T(k,\mu,z)&=\frac{\lambda^2\,T_{\rm sys}^2}{A_e\,t(k_\perp)}\;r^2(z)\;y(z),\\
y(z)&\equiv\frac{\dd r}{\dd \nu}=\frac{c(1+z)^2}{H(z)\,\nu_{21}},
\end{aligned}
\label{eq:NT}
\end{equation}
where $r(z)$ is the comoving distance and $A_e$ is the effective collecting area of a dish.
The time spent on a transverse mode is determined by the baseline density $\rho(U)$ through
\begin{equation}
t(k_\perp)=\frac{T_0\,N_{\rm ant}(N_{\rm ant}-1)\,A_e\,\rho(U)}{2\lambda^2},
~~
U=\frac{r(z)\,k_\perp}{2\pi},
\label{eq:tk}
\end{equation}
where $T_0$ is the observing time per pointing and $N_{\rm ant}$ is the number of dishes.
The variance in a $(k,\mu)$ bin is then approximated by
\begin{equation}
\delta P_{HI}(k,\mu,z)=
\frac{P_{HI}(k,\mu,z)+N_T(k,\mu,z)}{\sqrt{N_c(k,\mu,z)}}\,
\frac{1}{\sqrt{N_{\rm p}}},
\label{eq:sigmaP}
\end{equation}
with $N_{\rm p}$ being the number of independent pointings and
\begin{equation}
N_c(k,\mu,z)=
\frac{2\pi k^2\,\Delta k\,\Delta\mu}{(2\pi)^3}\,
r^2(z)\,y(z)\,B\,
\left(\frac{\lambda^2}{A_e}\right),
\label{eq:Nc}
\end{equation}
where $B$ is the frequency bandwidth and $(\Delta k,\Delta\mu)$ specify the bin widths; we use logarithmic binning with $\Delta k/k=1/6$.

\begin{figure}
\centering
\includegraphics[width=\columnwidth]{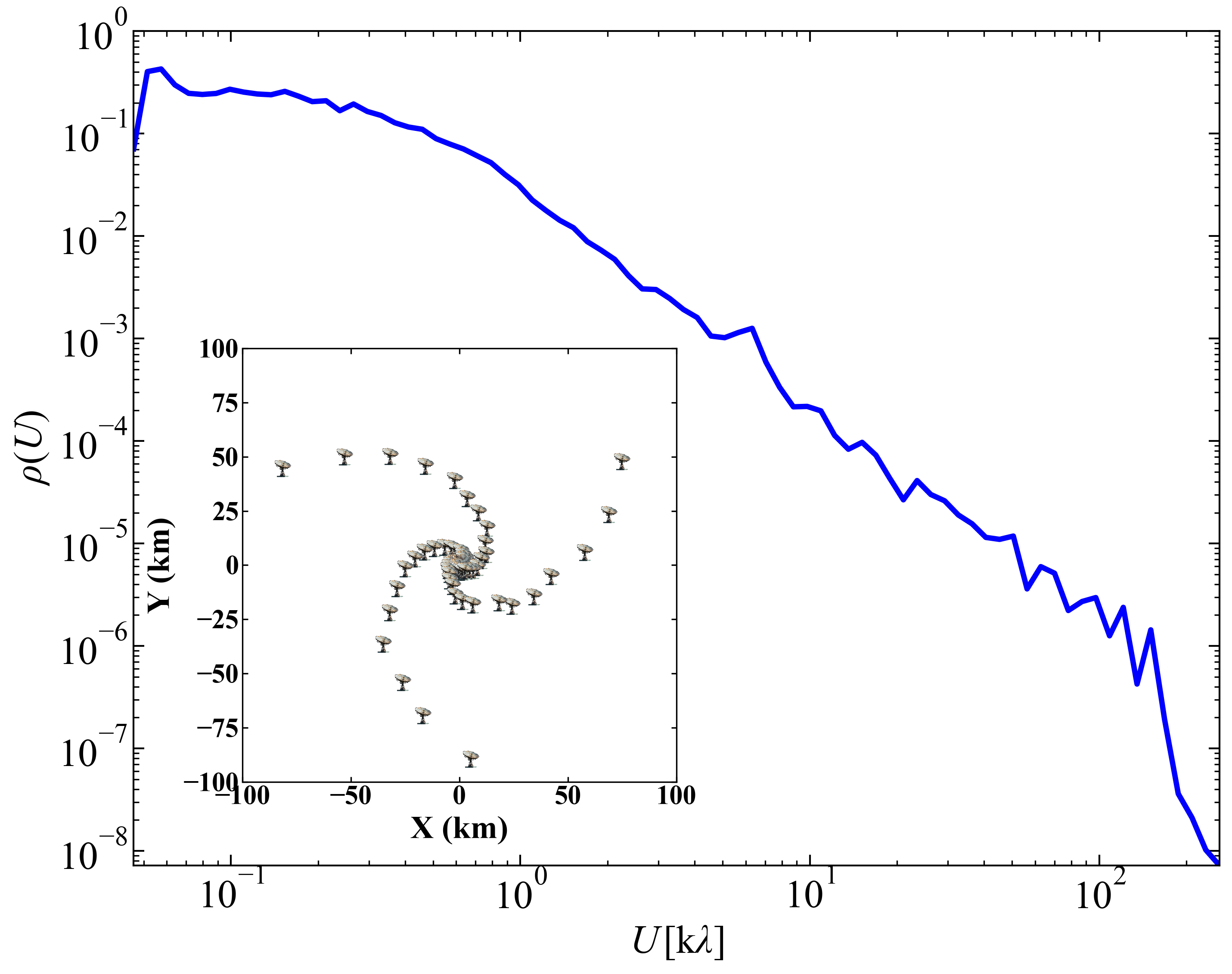}
\caption{Baseline distribution function $\rho(U)$ for the adopted SKA1-MID array configuration. The inset shows the actual antenna locations in the $X$--$Y$ plane.}
\label{fig:baseline}
\end{figure}

For the mock survey, we adopt an SKA1-MID-like interferometric configuration with
$N_{\rm ant}=197$ dishes of diameter $D=15~{\rm m}$ and aperture efficiency $\eta=0.7$
(so that $A_e=\eta\,\pi(D/2)^2$),
 a system temperature $T_{\rm sys}=60~{\rm K}$, and a bandwidth $B=128~{\rm MHz}$. 
We adopt the antenna locations of the SKA1-MID-like array radio interferometer\footnote{\url{https://www.skao.int/en}}. Fig.  \ref{fig:baseline} shows the normalized baseline distribution function $\rho(U)$ for this array.

We take a total observing time $T_{\rm tot}=4000~{\rm hr}$ split into $N_{\rm p}=10$ independent pointings (i.e.\ $T_0=400~{\rm hr}$ per pointing).
Our optimistic baseline 21\,cm forecast is evaluated at $z\simeq 1.75$ (corresponding to $\nu_{\rm obs}\simeq 517~{\rm MHz}$). This choice is well matched to the SKA1-MID observational configuration, since the corresponding frequency lies inside Band-1, where post-reionization HI intensity mapping can be carried out. This choice also provides comparatively good sensitivity for the adopted SKA1-MID-like interferometer. In particular, it is reported that the high-SNR region in the $(k_\parallel,k_\perp)$ plane is wider at $z=1.75$ compared to other redshifts around this epoch \citep{mohit}. Finally, this redshift lies safely within the calibrated redshift range of the emulator used for the DDM power-spectrum suppression \citep{Peters}. Thus, the adopted redshift is chosen to remain compatible with both the instrumental setup and the redshift range over which the power spectrum used in our modeling is valid. The mock 21\,cm modes entering the parameter analysis were selected using a fixed sensitivity criterion that favours modes with high signal-to-noise ratios and appreciable DDM-induced power suppression. The SNR maps and DDM detectability analysis at $z=1.75$ show that the scales carrying the relevant sensitivity lie comfortably below the calibrated DMemu/TBDemu limit, $k<6\,h\,{\rm Mpc}^{-1}$ \citep{mohit}. The $k$ values used in the present mock data therefore lie well inside the calibrated emulator domain. Applying an explicit $k<6\,h\,{\rm Mpc}^{-1}$ restriction does not remove any of the $k$-modes used in the mock data and leaves the posterior constraints and reconstructed quantities unchanged. The constant-suppression extrapolation beyond the calibrated emulator range is therefore not used.

\begin{table}
\centering
\resizebox{\columnwidth}{!}{%
\begin{tabular}{cccccc}
\hline
\hline
$N_{\mathrm{ant}}$ & Aperture efficiency & $D_{\mathrm{dish}}$ & $T = T_{0}\,N_{\mathrm{point}}$ & $T_{\mathrm{sys}}$ & $B$ \\
\hline
197 & 0.7 & 15\,m & 4000\,hrs & 60\,K & 128\,MHz \\
\hline
\hline
\end{tabular}%
}
\caption{The telescope specifications and observational parameters used in our analysis.}
\label{tab:telescope-params}
\end{table}

A few caveats should be kept in mind when interpreting the optimistic mock 21\,cm forecast used here. First, the adopted HI bias is taken from simulations calibrated in a stable cold dark matter scenario. In a true DDM cosmology, the suppression of small-scale matter power would reduce the abundance of low-mass haloes and shift the HI content toward larger haloes, which would generally enhance the small-scale HI bias relative to the CDM case. Our use of the CDM-based bias should therefore be regarded as a pragmatic approximation, and it may underestimate the small-scale 21\,cm signal in a DDM universe \citep{Sarkar_2016,mohit}. A full calibration of the HI bias in the DDM case would require dedicated $N$-body or hydrodynamical simulations and lies beyond the scope of the present work.

Second, we have not included an explicit shot-noise contribution in the 21\,cm power spectrum. This is consistent with the standard intensity-mapping assumption that the effective number density of DLA sources is sufficiently large for the clustering term to dominate on the scales of interest \citep{shotnoise}. Nevertheless, if the effective source density were lower than assumed, the shot-noise term would add an approximately scale-independent contribution to the auto-power spectrum and increase the total variance, thereby reducing the signal-to-noise ratio and weakening the forecasted parameter constraints.

Finally, we do not include foreground contamination in the present analysis. This should be viewed as an optimistic assumption. Galactic synchrotron emission, free-free emission, and extragalactic emission from radio point sources \citep{di2002radio,shaver1999can,2011MNRAS.418.2584G} are many orders of magnitude larger than the cosmological 21\,cm signal and require substantial foreground cleaning before the signal can be recovered. 

If the interferometer were perfectly achromatic, the smooth foregrounds would indeed remain confined near
$k_{\parallel} \approx0$. 
The foreground wedge appears because the interferometer is chromatic: its frequency-dependent response mixes angular structure ($k_\perp$)  into spectral structure ($k_\parallel$). Thus, although foregrounds intrinsically have power only near 
$k_{\parallel}\approx 0$, the instrument leaks this power to higher $k_{\parallel}$, with the amount of leakage increasing for longer baselines (larger $k_\perp$).  This chromatic mode mixing creates the characteristic wedge-shaped contaminated region in the $(k_\parallel, k_\perp)$ space, known as the foreground wedge. 
This region is roughly given by 
\begin{equation}
    k_\parallel  \leq \frac{H(z) D_M(z) \sin \theta_o }{c(1+z)} ~ k_\perp,
\end{equation}
where $\theta_o$ is the field of view of the telescope. A conservative analysis would require removal of all modes in this region, leading to the loss of a significant number of Fourier modes \citep{pober2013opening, pober2014next, liu2014epoch, dillon2015empirical, pal2021demonstrating}. 
In the context of the $\Lambda$DDM model, a simpler Fisher-matrix analysis has shown that the removed modes often lie in a region of $(k_\parallel,k_\perp)$ space where, due to the baseline coverage of the SKA1-MID instrument, the signal-to-noise ratio (in the absence of foregrounds)  is high. Thus, 
 removing these foreground-contaminated modes can substantially degrade the forecast (degradation by a factor of two to three of the constraints on the DDM parameters) \citep{mohit}.

 Additional observational issues, such as radio-frequency interference and bandpass calibration errors, can further reduce the effective sensitivity. For this reason, the joint DESI DR2 BAO + mock $21$\,cm constraints presented here should be interpreted as a proof-of-concept forecast under simplified observational assumptions rather than as a fully realistic survey prediction.

\section{Results and discussion}
\label{sec:results}
\begin{table}
\setlength{\tabcolsep}{4.5pt}
\renewcommand{\arraystretch}{1.15}

\begin{tabular}{l R@{$.$}L R@{$.$}L}
\hline\hline
Parameter & \multicolumn{2}{c}{DESI BAO} & \multicolumn{2}{c}{DESI BAO + mock $21$\,cm} \\
\hline\hline
$\log_{10}\epsilon$        & -1 & 268^{+0.674}_{-0.694} & ~~~~~~~~~-1 & 9786^{+0.134}_{-0.133} \\
\hline
$\log_{10}(\tau/{\rm yr})$ & 9 & 454^{+1.026}_{-1.001} & 10 & 4695^{+0.150}_{-0.146} \\
\hline
$h$                        & 0 & 674^{+0.011}_{-0.011} & 0 & 6730^{+0.011}_{-0.011} \\
\hline
$\Omega_{\rm dm}$          & 0 & 249^{+0.033}_{-0.034} & 0 & 2648^{+0.017}_{-0.018} \\
\hline
$a_1$                      & 0 & 775^{+0.291}_{-0.272} & 0 & 7648^{+0.282}_{-0.265} \\
\hline
$b_1$                      & 0 & 753^{+0.048}_{-0.053} & 0 & 7599^{+0.044}_{-0.047} \\
\hline
$c_1$                      & -0 & 423^{+0.116}_{-0.116} & -0 & 4325^{+0.107}_{-0.107} \\
\hline\hline
\end{tabular}
\centering

\caption{Posterior constraints on the semi-cosmographic parameters. The first column reports the constraints obtained using DESI DR2 BAO data. The second column reports the forecast obtained by combining the $21$\,cm intensity-mapping power spectrum at $z=1.75$ with the DESI DR2 BAO data.}

\label{tab:constraints}
\end{table}

\begin{figure*}
\centering
\includegraphics[width=0.49\textwidth]{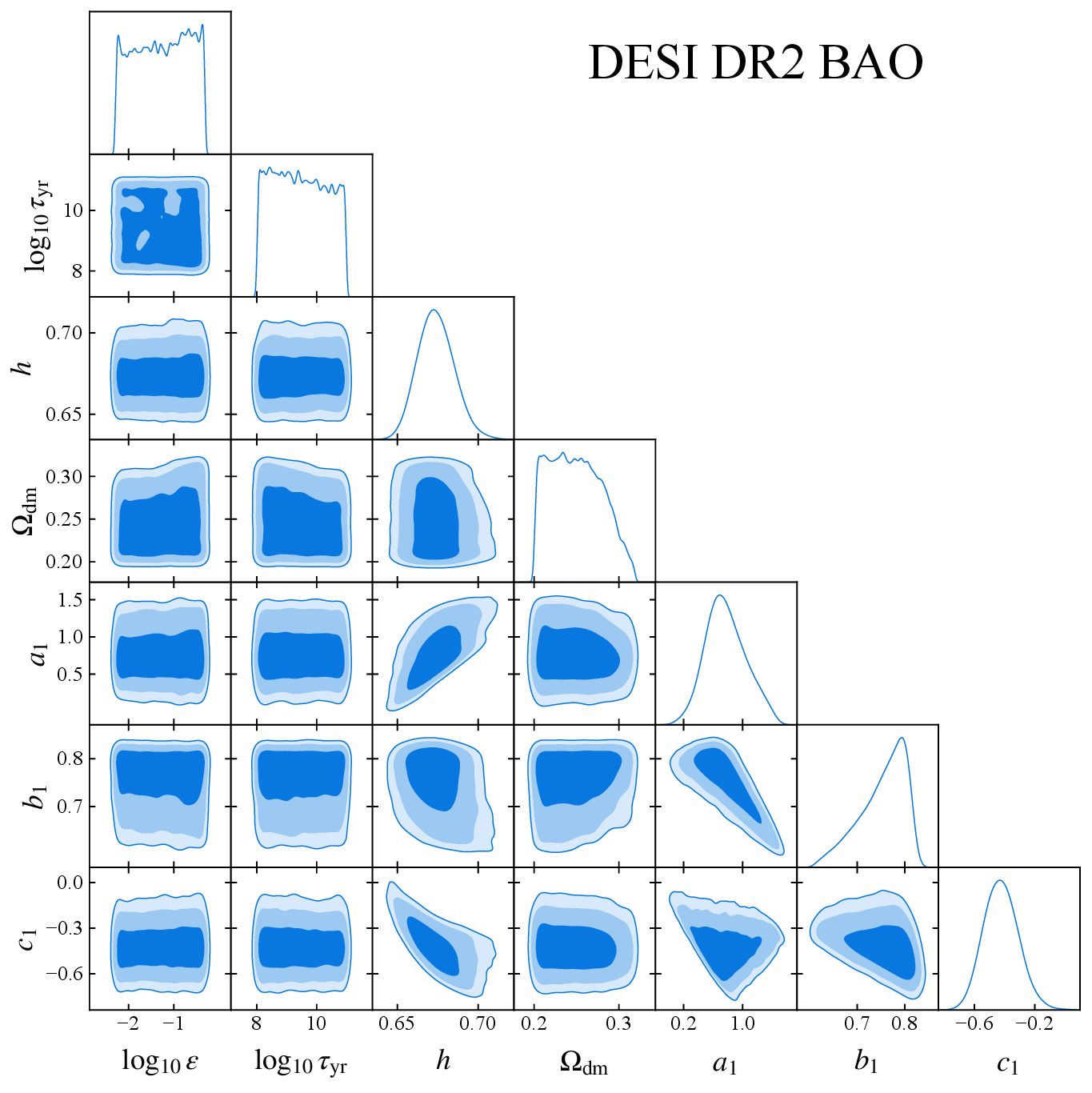}
\includegraphics[width=0.49\textwidth]{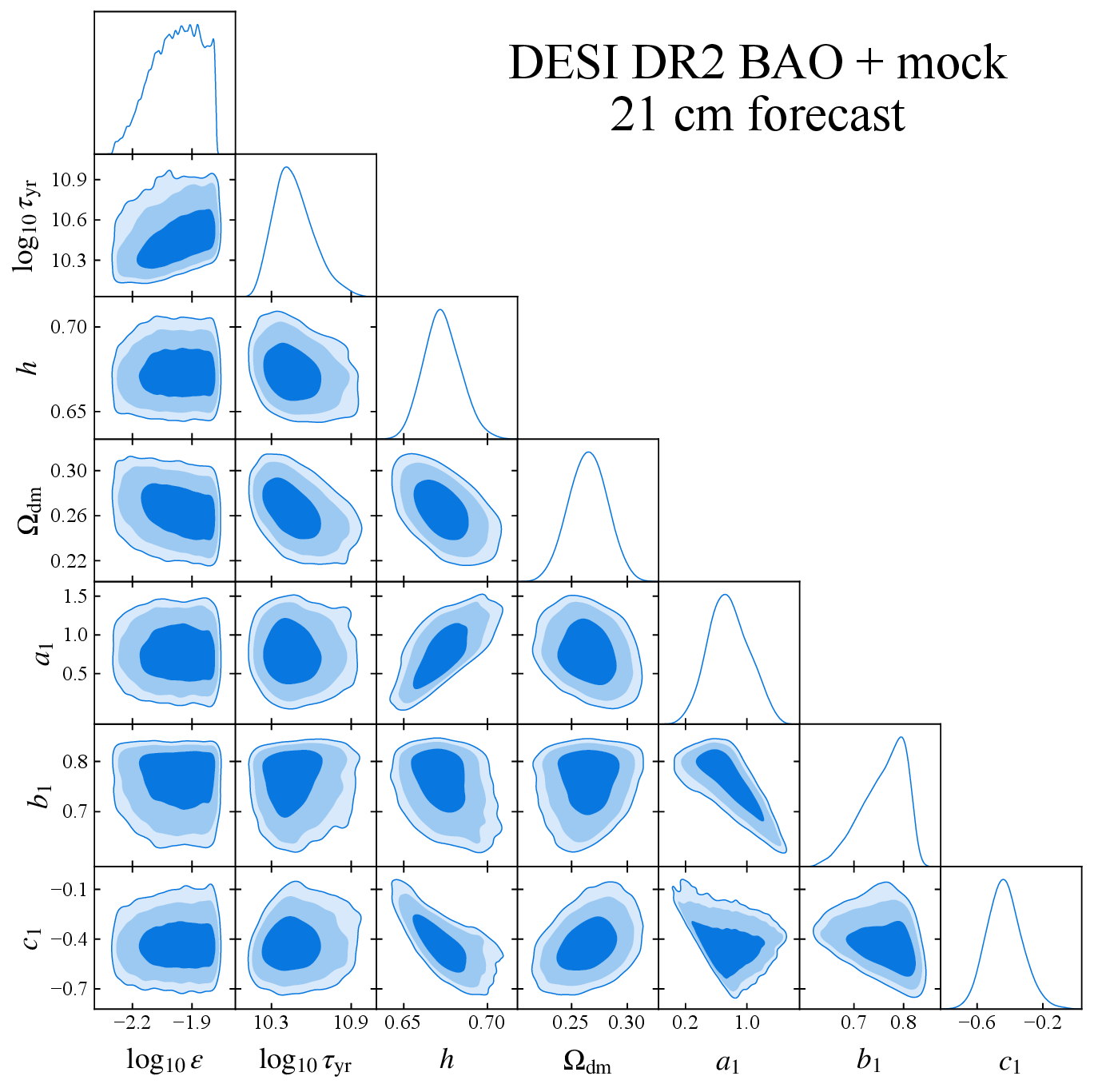}

\caption{Marginalized posterior distributions for the semi-cosmographic parameters.
\textit{Left:} Data -- DESI DR2 BAO only.
\textit{Right:} Forecast -- $21$\,cm power spectrum at $z=1.75$ with DESI DR2 BAO.
Contours show the $68\%$, $95\%$, and $99.7\%$ credible regions.}

\label{fig:triangle}
\end{figure*}

Fig.~\ref{fig:triangle} and Table~\ref{tab:constraints} summarize the MCMC posterior constraints on the Pad\'e and two-body DDM parameters. We first consider the DESI DR2 BAO data alone and then the joint constraints obtained by combining this data with the mock 21\,cm intensity-mapping power spectrum. Consequently, the joint posterior has support only within the calibrated emulator parameter domain. The wavenumbers used are well below the calibrated upper limit of $6\,h\,{\rm Mpc}^{-1}$, and no extrapolated suppression is used.

Using DESI DR2 BAO alone, we find that the Pad\'e parameters $(a_1,b_1,c_1)$ and the background parameters $(h,\Omega_{\rm dm})$ are constrained at the few-percent level.
This is expected, since BAO is mainly a geometric probe and is therefore very sensitive to the late-time expansion history through the distance combinations $\widetilde{D}_M$, $\widetilde{D}_H$, and $\widetilde{D}_V$.
In contrast, when only BAO data are used, the decay parameters $(\epsilon,\tau)$ remain only weakly constrained.
The reason is that BAO distances are sensitive mainly to the smooth background evolution, and in our framework, that evolution is already described by a flexible Padé form.
As a result, a range of different combinations of Pad\'e parameters and decay parameters can reproduce nearly the same distance-redshift relation.
In this sense, the decay effects in the DESI DR2 BAO only analysis are partly absorbed into the freedom of the Pad\'e expansion history, leaving a broad allowed region in the $(\epsilon,\tau)$ plane.

The situation changes when the mock $21$\,cm power spectrum is included.
Unlike BAO, the $21$\,cm signal is sensitive not only to the background expansion but also to the growth rate and scale dependence of matter clustering.
This makes it much more directly sensitive to the physical effects of decaying dark matter.
The decay parameters $(\epsilon,\tau)$ determine how much of the dark matter has decayed by a given epoch and how strongly the massive daughter is kicked at its time of production.
These two effects change the shape and overall suppression of the matter power spectrum.
By contrast, the Pad\'e parameters mainly alter the smooth background evolution through $H(z)$.
Therefore, once the $21$\,cm power spectrum is added, the data can distinguish between a change in the expansion history and a genuine decay-induced suppression in clustering.
Thus, the joint DESI DR2 BAO + mock $21$\,cm analysis is able to pin down the decay sector much more effectively than DESI DR2 BAO alone.
Quantitatively, the addition of the mock $21$\,cm likelihood sharpens the decay constraints to
$\log_{10}\epsilon=-1.9786^{+0.134}_{-0.133}$ and
$\log_{10}(\tau/{\rm yr})=10.4695^{+0.150}_{-0.146}$,
which correspond to $\epsilon\simeq 1.1\%$ and $\tau\simeq 29~{\rm Gyr}$ at $68\%$ credibility.
The shrinkage of the allowed region compared with the DESI DR2 BAO only case is substantial and clearly shows that the clustering information is doing the main work in constraining the DDM parameters.

Our analysis uses a fixed value of $r_d$ adopted from the Planck CMB constraints within the standard $\Lambda$CDM model. In a general DDM model with a relativistic daughter component, this approximation requires a justification. A positive equation of state for the massive daughter, $w_2(z) > 0$, makes that component redshift faster than ordinary pressureless matter, while the massless daughter contributes to the radiation sector. Therefore, the DDM sector can, in principle, modify the matter-radiation equality epoch, the pre-recombination expansion rate, and the sound horizon $r_d$.

For this purpose, each posterior sample was compared with a corresponding no-decay case.  This gives a direct estimate of how strongly the decay process, characterized by the DDM parameters, affects the early-Universe cosmology.  The sound horizon is evaluated using
\begin{equation} 
r_d=\int_{z_d}^{\infty}\frac{c_s(z)}{H(z)}\,dz,\qquad
c_s(z)=\frac{c}{\sqrt{3[1+R_b(z)]}},
\end{equation}
where $R_b(z)=3\rho_b/(4\rho_\gamma)$ is the baryon-to-photon density ratio, and we take the drag redshift to be $z_d=1060$. 
We also quantify the fraction of the unstable component that has decayed by a cosmic time $t(z)$. This is estimated as $ f_{\rm dec}(z)=1-\exp[-t(z)/\tau]$.

We then quantified three fractional shifts: $\Delta H/H$, which measures the change in the pre-recombination expansion rate; $\Delta z_{\rm eq}/z_{\rm eq}$, which measures the change in the matter-radiation equality redshift; and $\Delta r_d/r_d$, which measures the fractional change in the sound horizon at the drag-epoch. We define 
\begin{eqnarray}
\left|\frac{\Delta H}{H}\right|&=\left|\frac{H_{\rm DDM}-H_{\rm ref}}{H_{\rm ref}}\right|,\\
\left|\frac{\Delta z_{\rm eq}}{z_{\rm eq}}\right|&=\left|\frac{z_{{\rm eq},{\rm DDM}}-z_{{\rm eq},{\rm ref}}}{z_{{\rm eq},{\rm ref}}}\right|,
\\
\left|\frac{\Delta r_d}{r_d}\right|&=\left|\frac{r_{d,{\rm DDM}}-r_{d,{\rm ref}}}{r_{d,{\rm ref}}}\right|,
\end{eqnarray}
where the subscript ``ref'' denotes the no-decay scenario. 
These quantities test whether the DDM posterior region changes the early-time expansion history enough to invalidate the fixed-$r_d$ BAO calibration.
The posterior samples from both the DESI DR2 BAO only and DESI DR2 BAO + mock $21$\,cm chains are applied to these, and their maximum values are recorded. The results are summarized in
Table \ref{tab:rd_consistency}. 
\begin{table*}
\centering
\begin{tabular}{lcccc}
\hline
\hline
Posterior &
~~~$\max f_{\rm dec}$ &
~~$\max |\Delta H/H|$ &
~~$\max |\Delta z_{\rm eq}/z_{\rm eq}|$ &
~~$\max |\Delta r_d/r_d|$ \\
\hline
DESI DR2 & ~~~$4.07\times10^{-3}$ & ~~$4.33\times10^{-4}$ & ~~$3.88\times10^{-4}$ & ~~$1.15\times10^{-4}$ \\
DESI DR2 + mock $21$\,cm & ~~~$2.71\times10^{-5}$ & ~~$1.04\times10^{-7}$ & ~~$1.80\times10^{-6}$ & ~~$2.75\times10^{-8}$ \\
\hline
\hline
\end{tabular}
\caption{Maximum early-time consistency diagnostics over the posterior samples.}
\label{tab:rd_consistency}
\end{table*}
The results show that almost no decay occurs before the drag epoch, and the sound horizon is practically unchanged.

The largest sampled sound-horizon shift is below the relative uncertainty of the adopted CMB sound horizon, $0.23/147.21\simeq1.56\times10^{-3}$. Thus, the fixed-$r_d$ treatment is self-consistent within the allowed posterior regions.

It is useful to compare our inferred DDM parameters with representative values reported in earlier studies.
As reported in Abell\'an et al.~\citep{abellan2021linear}, several of the tabulated DDM fits include central values and $1\sigma$ uncertainties for \(\log_{10}\epsilon\) and \(\log_{10}\tau\), allowing a direct comparison with our MCMC posteriors.
For these parameter values, we quantify the tension in each parameter of interest.
The resulting tensions are listed in Table~\ref{tab:tension_abellan}.

The work by Fu{\ss} \& Garny~\citep{ddmdata} reports the best-fit points rather than independent  $1\sigma$ ranges for \((\epsilon,\tau)\).
For this reason, we do not quote a strict Gaussian tension relative to their results.
Instead, we report in Table~\ref{tab:offset_fuss} the offset of their best-fit points measured in units of the corresponding posterior widths.
This provides a simple measure of proximity to the best-fit values reported in the literature without assigning to these offsets the statistical meaning of a formal tension.

The comparison shows that the DESI DR2 BAO only constraints are broadly compatible with the earlier constraints on the  DDM parameters, which is expected given the broad decay-sector posteriors in the DESI DR2 BAO only fit.
In the joint DESI DR2 BAO + mock $21$\,cm case, the inferred value of \(\epsilon\) remains in the same low-\(\epsilon\) regime favoured by previous work, while the preferred lifetime shifts toward somewhat shorter values.
Among the results reported in Abell\'an et al., the closest agreement is found with their Planck full/lite and \(S_8\)-motivated solutions, for which both decay parameters remain within about \(1\sigma\).
In the Fu{\ss} \& Garny comparison, our joint result remains close to their BestFit2 point, while it is far from BestFit1, which lies at much smaller \(\epsilon\) and a much longer lifetime.
Overall, our results support the same broad low-\(\epsilon\), long-\(\tau\) DDM regime identified in the literature, but indicate a preference for a somewhat shorter lifetime once the mock $21$\,cm clustering information is included.

\begin{table*}
\centering
\small
\begin{tabular}{llccccc}
\hline
\hline
 Case & $\epsilon_{\rm lit}$ & $\tau_{\rm lit}$ [Gyr] &
$T_{\epsilon}^{\rm DATA}$ & $T_{\tau}^{\rm DATA}$ &
$T_{\epsilon}^{\rm FORECAST}$ & $T_{\tau}^{\rm FORECAST}$ \\
\hline
\hline
BAO+SNIa+Planck & $2.04\times 10^{-3}$ & --    & 2.05 & --   & 5.35 & --   \\
KiDS+BOSS+2dFLens & $5.25\times 10^{-3}$ & 77.6  & 1.46 & 1.40 & 2.27 & 2.80 \\
DES & $<7.24\times 10^{-3}$            & 141.3 & --   & 1.65 & --   & 4.54 \\
KiDS+Viking+DES & $5.89\times 10^{-3}$ & 41.7  & 1.39 & 1.14 & 1.89 & 1.00 \\
$A_{\rm lens}$ marginalized & --      & 125.9 & --   & 1.60 & --   & 4.20 \\
fixed $\epsilon=0.05$      & $5.00\times 10^{-2}$ & 524.8 & -- & 2.21 & -- & 8.33 \\
SPTpol without $S_8$ prior & --      & 239.9 & --   & 1.88 & --   & 6.07 \\
SPTpol with $S_8$ prior    & --      & 177.8 & --   & 1.75 & --   & 5.20 \\
ACTPol without $S_8$ prior & --      & 0.120 & --   & 1.37 & --   & 16.39 \\
ACTPol with $S_8$ prior    & $4.57\times 10^{-3}$ & 0.0372 & 1.55 & 1.88 & 2.72 & 19.87 \\
Planck full                & $5.50\times 10^{-3}$ & 69.2  & 1.43 & 1.35 & 2.11 & 2.47 \\
Planck lite                & $5.25\times 10^{-3}$ & 77.6  & 1.46 & 1.40 & 2.27 & 2.80 \\
\hline
\hline
\end{tabular}

\caption{Parameter tensions between our DDM constraints and the values reported in~\citep{abellan2021linear}. The columns marked DATA refer to the DESI DR2 BAO current-data fit, while the columns marked FORECAST refer to the joint DESI DR2 BAO + mock 21\,cm forecast. The tensions $T_{\epsilon}$ and $T_\tau$ are computed for the sampled variables \(\log_{10}\epsilon\) and \(\log_{10}(\tau/{\rm yr})\), using the \(1\sigma\) uncertainties.}
\label{tab:tension_abellan}
\end{table*}
\begin{table*}
\centering
\small
\begin{tabular}{lcccccc}
\hline
\hline
Case & $\epsilon_{\rm lit}$ & $\tau_{\rm lit}$ [Gyr] &
$\Delta_{\epsilon}^{\rm DATA}$ & $\Delta_{\tau}^{\rm DATA}$ &
$\Delta_{\epsilon}^{\rm FORECAST}$ & $\Delta_{\tau}^{\rm FORECAST}$ \\
\hline
BestFit1 (Planck+BAO+FS) & $1.45\times 10^{-4}$ & 955.0 & 3.71 & 2.46 & 13.99 & 10.07 \\
BestFit2 (+KiDS $S_8$) & $1.20\times 10^{-2}$ & 120.2 & 0.94 & 1.58 & 0.43 & 4.07 \\
\hline
\hline
\end{tabular}
\caption{Offsets between our inferred DDM parameters and the best-fit values reported by~\citep{ddmdata}. The columns marked DATA refer to the DESI DR2 BAO current-data fit, while the columns marked FORECAST refer to the joint DESI DR2 BAO + mock 21\,cm forecast. Since Fu{\ss} \& Garny provide best-fit values rather than independent symmetric \(1\sigma\) constraints on \((\epsilon,\tau)\), the quantities \(\Delta_{\epsilon}\) and \(\Delta_{\tau}\) are reported in units of the \(1\sigma\) posterior width from our analysis and should be interpreted as best-fit offsets rather than formal tensions.}

\label{tab:offset_fuss}
\end{table*}

\begin{figure*}
\centering
\includegraphics[width=0.49\textwidth]{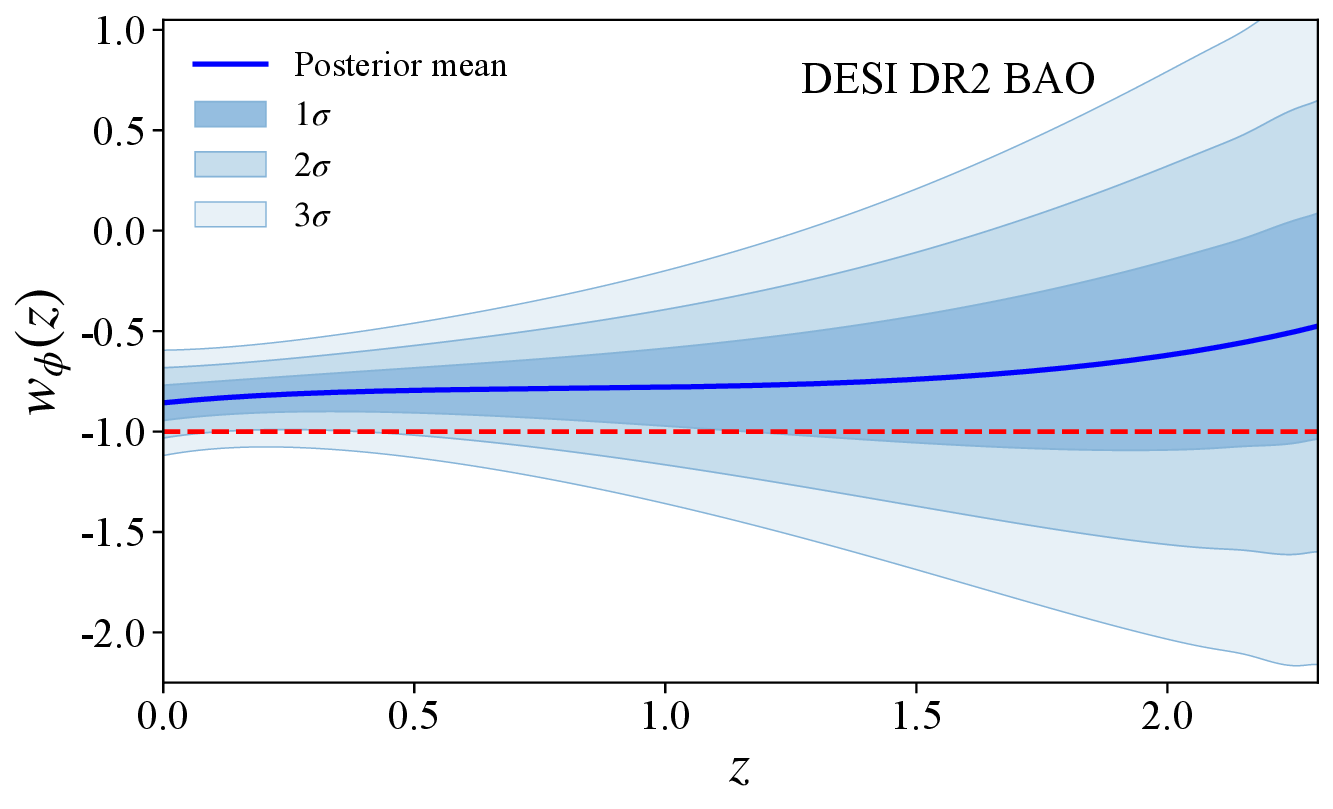}
\includegraphics[width=0.50\textwidth]{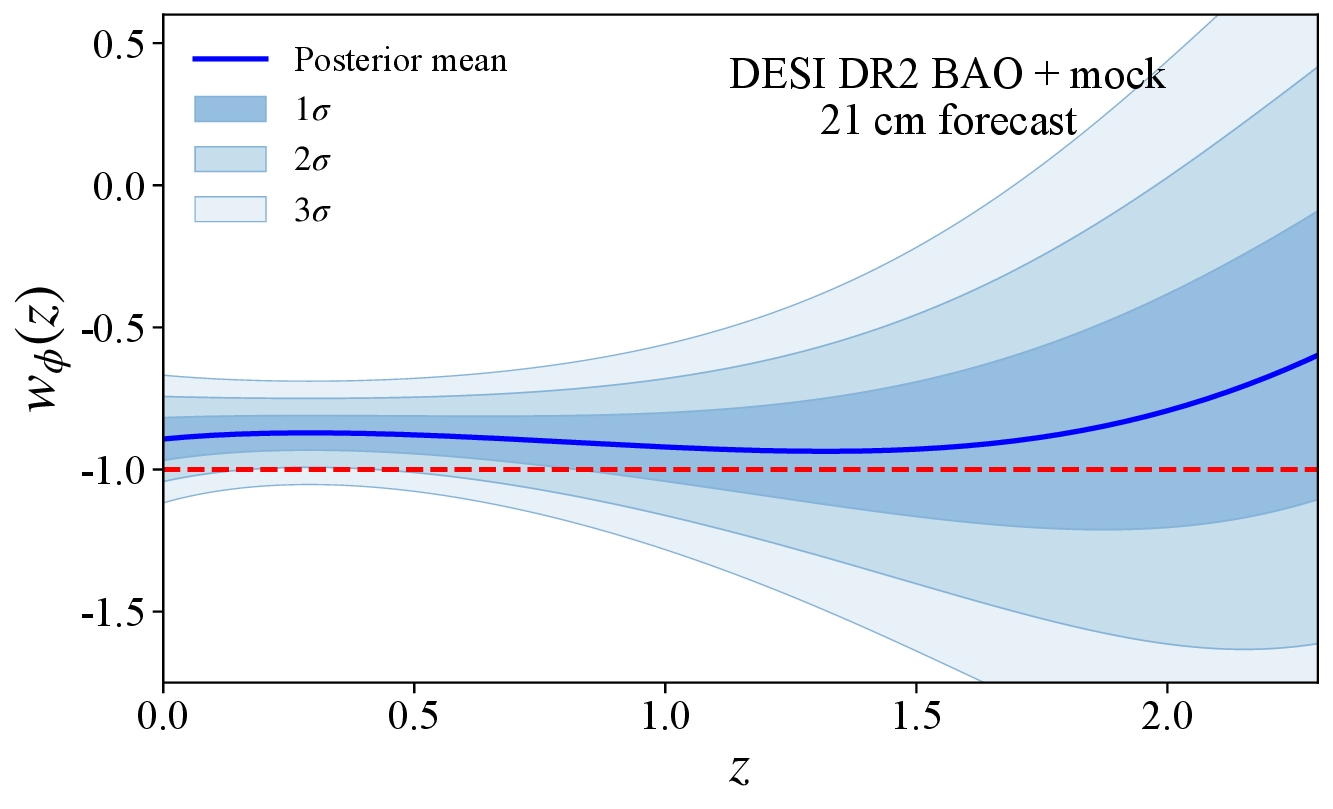}

\caption{Residual dark-energy reconstruction of the equation of state $w_\phi(z)$ inferred from the semi-cosmographic expansion history constructed using the Pad\'e approximation and the two-body DDM model. 
\textit{Left:} Data -- DESI DR2 BAO only.
\textit{Right:} Forecast -- $21$\,cm power spectrum with DESI DR2 BAO.
The solid curve shows the posterior mean, while shaded bands indicate $\pm1\sigma$, $\pm2\sigma$, and $\pm3\sigma$ ranges.
The dashed line marks $\LCDM$ ($w_\phi=-1$).}
\label{fig:wphi}
\end{figure*}

The reconstructed $w_\phi(z)$ is obtained by subtracting the contributions of DDM, baryons, and radiation from the Pad\'e-reconstructed expansion history under the assumptions of flat FLRW cosmology and the standard Friedmann equations. Thus, it must be interpreted only as an effective residual dark-energy equation of state rather than an independently parameterized dark energy model. The Pad\'e approximation provides a flexible representation of the total background expansion history through $D_L(z)$ and the corresponding $H(z)$ derived from it, but it does not distinguish between the individual physical components contributing to the expansion. In particular, the Pad\'e ansatz contains no explicit decomposition into baryonic matter, dark matter, radiation, and dark energy. Any smooth feature in the reconstructed background that is not accounted for by the assumed matter sector and radiation is therefore attributed to the effective residual dark-energy component.
The equation of state of the effective reconstructed dark energy, $w_\phi(z)$, is therefore not an independently fitted quantity. Accordingly, $w_\phi$ is evaluated for every MCMC sample, so that each realization inherits the full posterior covariance of the fitted Padé, DDM, and cosmological parameters. No additional fit of $w_\phi(z)$ is performed, and no extra free parameters are introduced.

Since the same reconstructed expansion history determines both the geometry-based observables and the residual dark-energy properties, the uncertainties in $w_\phi(z)$ are intrinsically correlated with those in the reconstructed distances and $H(z)$ (not shown in this work). These correlations are automatically propagated through the MCMC sampling. Consequently, the uncertainty bands on $w_\phi(z)$ should not be interpreted as those of an independent dark energy model, but rather as the conditional uncertainty of the effective reconstructed dark energy within the adopted Padé parameterization, flat FLRW geometry, and DDM framework. Because $w_\phi(z)$ is derived only after the likelihood evaluation and is not used as an additional observable in the fit, there is no double counting of information; however, its reconstruction and associated uncertainties remain conditional on the adopted ansatz and naturally inherit the parameter degeneracies of the underlying model.

We must also note that different parameterizations of the background expansion or different assumptions about the matter sector could lead to quantitatively different residual dark-energy reconstructions, even when they provide comparably good fits to the observational data.

For some parameter realizations encountered during MCMC sampling, the effective residual dark-energy equation of state $w_{\phi}^{\mathcal P}(z) \propto 1/ \Omega_{\phi}^{\mathcal P}$ diverges when $\Omega_{\phi}^{\mathcal P}(z)\rightarrow 0$. That is to say, the sum of the DDM, baryon, and radiation density fractions becomes equal to unity, indicating that no effective residual dark-energy component is present. Some parameter combinations can avoid divergence, but this cannot be guaranteed \textit{a priori} while running the MCMC chains. 

Therefore, to interpret this residual density $\Omega_\phi^{\mathcal P}(z)$ as a physical effective residual dark-energy density, we explicitly impose the physically motivated top-hat priors $H^{\mathcal P}(z)>0$ and $\Omega_\phi^{\mathcal P}(z)>0$ during MCMC sampling. We note that all accepted MCMC samples have no Pad\'e poles in the redshift interval used in our analysis.

Fig.~\ref{fig:wphi} shows the equation of state of the effective reconstructed dark energy. The left panel shows the results with  DESI only data, whereas the right panel shows the reconstruction from the joint DESI DR2 BAO + optimistic mock 21\,cm  analysis. In the DESI-only reconstruction, the $\Lambda$CDM line $w_\phi=-1$ lies well within the broad posterior support, showing that BAO distance information by itself is insufficient to distinguish a cosmological constant from mildly evolving effective reconstructed dark energy. The broad overlap with $w_\phi=-1$ reflects the fact that, in the absence of growth-sensitive information, a sizeable part of the uncertainty in the Pad\'e and DDM sectors propagates directly into the reconstructed $w_\phi(z)$.

In the joint forecast with  DESI DR2 BAO + mock $21$\,cm, the reconstructed band is visibly tighter, and the overlap with $w_\phi=-1$ is much smaller. This suggests that the clustering information is beginning to disfavor the region of parameter space closest to an exact cosmological constant and is instead mildly favoring solutions with $w_\phi(z)>-1$. Even so, the $\Lambda$CDM limit is not fully excluded by the reconstruction (especially at large redshifts), and therefore the result should be interpreted as a weak tension with a pure cosmological constant rather than as evidence against it. A more definitive statement would require either real $21$\,cm data or the inclusion of additional growth-sensitive probes.
In both the DESI DR2 BAO and joint DESI DR2 BAO + optimistic mock 21\,cm analyses, the cosmological constant value $w_\phi=-1$ remains consistent with the reconstructed bands over the full redshift range shown.
The posterior mean shows a mild trend toward $w_\phi>-1$ at $z\gtrsim 1.5$, but the uncertainties also increase significantly with redshift.
For this reason, we do not interpret this as evidence for a physical departure from a cosmological constant.
The main effect of adding the $21$\,cm information is to reduce the uncertainty in the DDM sector, which in turn reduces the uncertainty propagated into the reconstruction of $w_\phi(z)$, especially at low and intermediate redshifts.
At higher redshifts, however, the allowed range of $w_\phi(z)$ remains broad.

\begin{figure*}
\centering
\includegraphics[width=0.49\textwidth, height=5.0cm, keepaspectratio]{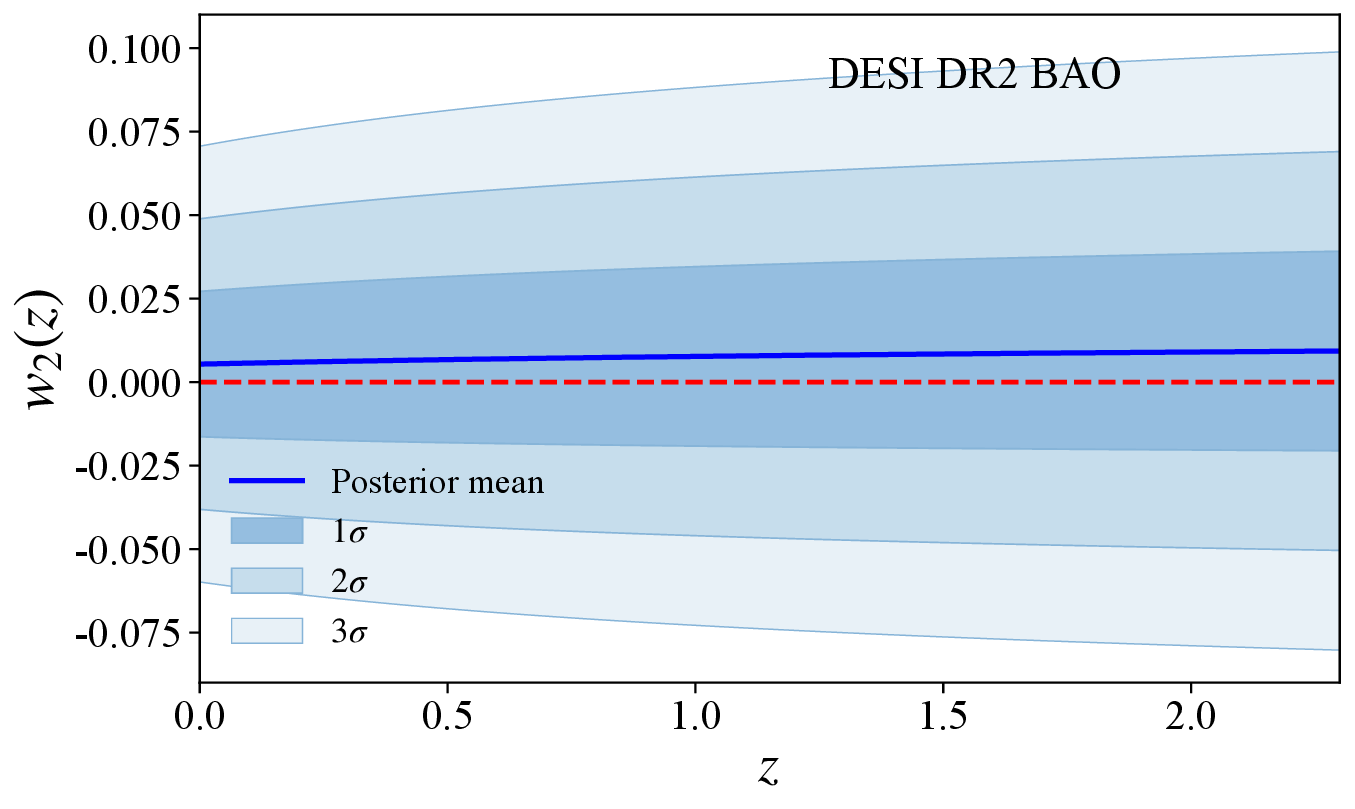}
\includegraphics[width=0.48\textwidth, height=5.0cm, keepaspectratio]{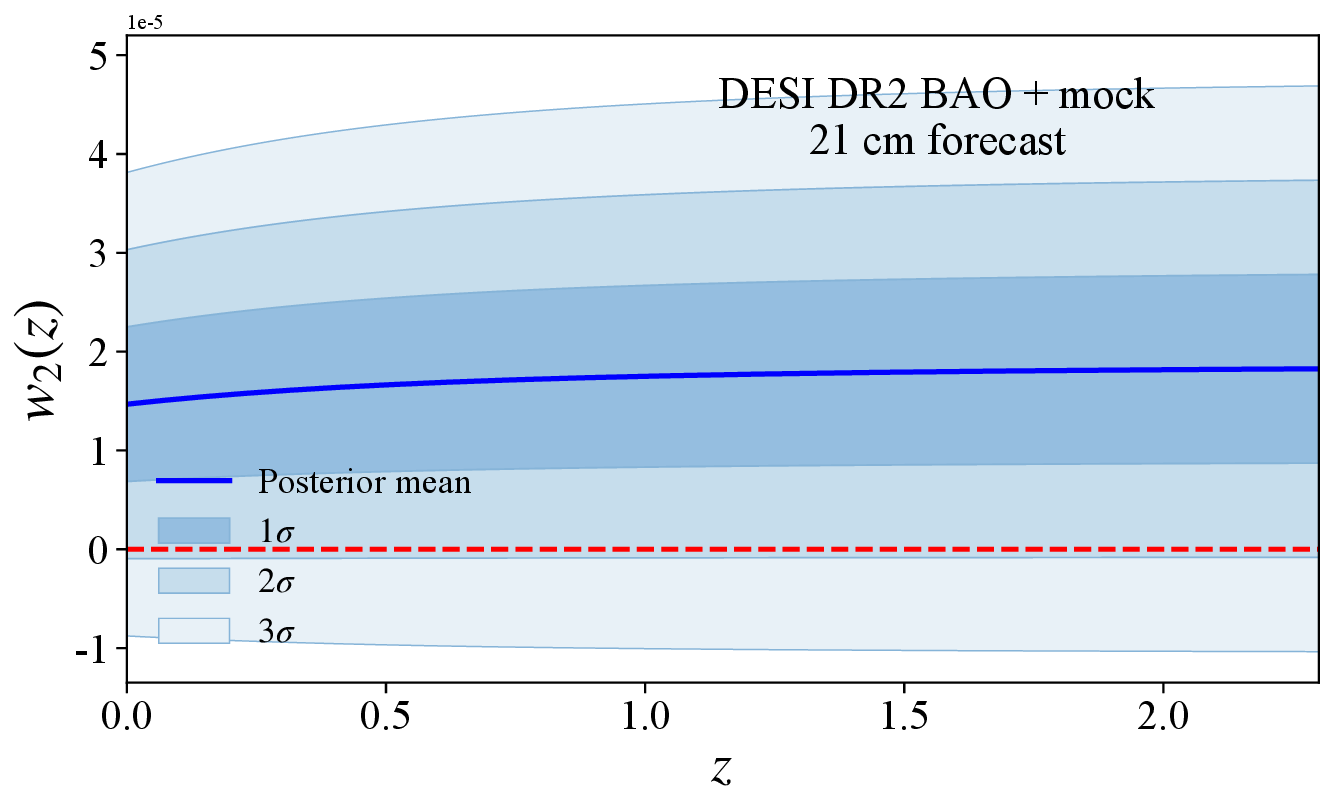}
\caption{Reconstruction of the equation of state $w_2(z)$ of the massive daughter particle.
\textit{Left:} Data -- DESI DR2 BAO only.
\textit{Right:} Forecast -- $21$\,cm power spectrum with DESI DR2 BAO.
The solid curve shows the posterior mean, while shaded bands indicate $\pm1\sigma$, $\pm2\sigma$, and $\pm3\sigma$ ranges. The dashed line marks the cold limit $w_2=0$.}
\label{fig:w2}
\end{figure*}

Fig.~\ref{fig:w2} shows the reconstructed equation of state of the massive daughter.
DESI DR2 BAO alone provides only weak information on $w_2(z)$, and the reconstructed band remains consistent with the cold limit within the uncertainty region.
This is again a direct consequence of the broad DESI DR2 BAO only posterior in the decay parameters, since $w_2(z)$ is fully determined once $(\epsilon,\tau)$ are fixed.
After including the mock $21$\,cm power spectrum, the posterior confines $w_2(z)$ to $\mathcal{O}(10^{-5})$ over $0\le z\le 2.3$.
This means that, in the region preferred by the joint DESI DR2 BAO + mock 21\,cm analysis, the massive daughter behaves effectively as a cold component for late-time structure formation.
The strong tightening of $w_2(z)$ provides a simple and intuitive way of seeing how the $21$\,cm data constrain the decay sector. Once the clustering suppression is measured, only a narrow range of decay histories remains allowed, and the behaviour of the massive daughter is forced to be extremely close to cold dark matter.

\section{Conclusions}
\label{sec:conclusion}

We have presented a hybrid phenomenological semi-cosmographic framework in which the late-time expansion history is described by a Pad\'e rational approximation for the luminosity distance. The effective reconstructed dark energy is described through its EoS within a setup defined by flat FLRW geometry, a Pad\'e expansion history, and a two-body decaying dark matter scenario.

Using DESI DR2 BAO alone, we find that the Pad\'e parameters and background densities are constrained at the few-percent level, but the decay parameters remain weakly constrained.
This is because BAO mainly probes the smooth background geometry, and the effect of decay on the expansion history can be partly absorbed by the flexibility of the Pad\'e parametrization.
When the mock $21$\,cm intensity-mapping power spectrum is added, the decay parameters become tightly constrained.
The reason is that the $21$\,cm power spectrum is directly sensitive to the suppression of matter clustering caused by the decay process, and this effect cannot be mimicked simply by changing the background expansion.
In this way, the addition of $21$\,cm information breaks the degeneracy between the Pad\'e parameters and the DDM parameters.

The equation of state of the effective reconstructed dark energy remains consistent with a cosmological constant within the present uncertainties. While the DESI DR2 BAO reconstruction leaves $\Lambda$CDM comfortably allowed, the joint DESI DR2 BAO + mock $21$\,cm case reduces the overlap with $w_\phi=-1$. This suggests that growth-sensitive information provides a more sensitive test than geometry-only information.
We also note that the reconstructed equation of state of the massive daughter is driven to very small values in the joint DESI DR2 BAO + mock 21\,cm analysis, showing that the daughter behaves effectively as a cold component at late times.
Our method, therefore, combines a flexible data-driven expansion history with a phenomenologically motivated dark matter physics, so the resulting residual dark-energy reconstruction is conditional on that hybrid setup. 
We note that for the mock 21\,cm intensity mapping, we have assumed a highly idealized observation. We emphasize that it is a proof-of-concept forecast . Foregrounds, systematics, and calibration errors would significantly degrade the error projections. However, poor constraints on the DDM parameters even in this ideal scenario imply that clustering information from other probes, such as weak lensing or Lyman-$\alpha$ forest, may be more useful. Cross-correlation power-spectrum studies may also prove to be useful. 

Overall, our results show that combining geometric probes with clustering information is essential for isolating the physical effects of decaying dark matter within a hybrid semi-cosmographic framework.

\begin{acknowledgments}
T.G.S acknowledges financial support from the ARG–MATRICS project ANRF\slash ARGM\slash 2025\slash 002607\slash TS.
The authors (MY) and (PC) acknowledge Birla Institute of Technology and Science, Pilani, Pilani Campus, Rajasthan, for financial support.
\end{acknowledgments}

\section*{Data Availability}
The DESI BAO measurements used in this work are publicly available from the DESI Collaboration releases.


\bibliographystyle{apsrev4-2}
\bibliography{mybib}

@STRING(Jan="January")

@STRING(Feb="February")

@STRING(Mar="March")

@STRING(Apr="April")

@STRING(May="May")

@STRING(Jun="June")

@STRING(Jul="July")

@STRING(Aug="August")

@STRING(Sep="September")

@STRING(Oct="October")

@STRING(Nov="November")

@STRING(Dec="December")

@string{mnras="Monthly Notices Of the Royal Astronomical Society"}

@string{apj="The Astrophysical Journal "}

@string{prd="Physical Review D"}

@article{turner2000dark,
  title={The dark side of the universe: from Zwicky to accelerated expansion},
  author={Turner, Michael S},
  journal={Physics Reports},
  volume={333},
  pages={619--635},
  year={2000},
  publisher={Elsevier}
}

@article{davis1985evolution,
  title={The evolution of large-scale structure in a universe dominated by cold dark matter},
  author={Davis, Marc and Efstathiou, George and Frenk, Carlos S and White, Simon DM},
  journal={Astrophysical Journal, Part 1 (ISSN 0004-637X), vol. 292, May 15, 1985, p. 371-394. Research supported by the Science and Engineering Research Council of England and NASA.},
  volume={292},
  pages={371--394},
  year={1985}
}

@article{ibarra2013indirect,
  author = "Ibarra, Alejandro and Tran, David and Weniger, Christoph",
    title = "{Indirect Searches for Decaying Dark Matter}",
    eprint = "1307.6434",
    archivePrefix = "arXiv",
    primaryClass = "hep-ph",
    doi = "10.1142/S0217751X13300408",
    journal = "Int. J. Mod. Phys. A",
    volume = "28",
    pages = "1330040",
    year = "2013"
}

@article{abellan2021linear,
  author = "Franco Abell\'an, Guillermo and Murgia, Riccardo and Poulin, Vivian",
    title = "{Linear cosmological constraints on two-body decaying dark matter scenarios and the S8 tension}",
    eprint = "2102.12498",
    archivePrefix = "arXiv",
    primaryClass = "astro-ph.CO",
    doi = "10.1103/PhysRevD.104.123533",
    journal = "Phys. Rev. D",
    volume = "104",
    number = "12",
    pages = "123533",
    year = "2021"
}

@article{Vattis,
  title = {Cosmological constraints on late-Universe decaying dark matter as a solution to the ${H}_{0}$ tension},
  author = {Clark, Steven J. and Vattis, Kyriakos and Koushiappas, Savvas M.},
  journal = {Phys. Rev. D},
  volume = {103},
  issue = {4},
  pages = {043014},
  numpages = {13},
  year = {2021},
  month = {Feb},
  publisher = {American Physical Society},
  doi = {10.1103/PhysRevD.103.043014},
  url = {https://link.aps.org/doi/10.1103/PhysRevD.103.043014}
}

@article{blackadder2016cosmological,
  author = {{Blackadder}, Gordon and {Koushiappas}, Savvas M.},
        title = "{Cosmological constraints to dark matter with two- and many-body decays}",
      journal = {Physical Review D},
         year = 2016,
        month = jan,
       volume = {93},
       number = {2},
          eid = {023510},
        pages = {023510},
          doi = {10.1103/PhysRevD.93.023510},
archivePrefix = {arXiv},
       eprint = {1510.06026},
 primaryClass = {astro-ph.CO},
       adsurl = {https://ui.adsabs.harvard.edu/abs/2016PhRvD..93b3510B}
}

@article{vattis2019dark,
    author = "Vattis, Kyriakos and Koushiappas, Savvas M. and Loeb, Abraham",
    title = "{Dark matter decaying in the late Universe can relieve the H0 tension}",
    eprint = "1903.06220",
    archivePrefix = "arXiv",
    primaryClass = "astro-ph.CO",
    doi = "10.1103/PhysRevD.99.121302",
    journal = "Phys. Rev. D",
    volume = "99",
    number = "12",
    pages = "121302",
    year = "2019"
}

@article{Planck2018,
   title={Planck 2018 results},
   volume={641},
   ISSN={1432-0746},
   url={http://dx.doi.org/10.1051/0004-6361/201833910},
   DOI={10.1051/0004-6361/201833910},
   journal={Astronomy and Astrophysics},
   publisher={EDP Sciences},
   author={Aghanim, N. and Akrami, Y. and Ashdown, M. and Aumont, J. and Baccigalupi, C. and Ballardini, M. and Banday, A. J. and Barreiro, R. B. and Bartolo, N. and et al.},
   year={2020},
   month={Sep},
   pages={6-6}
}

@ARTICLE{2011MNRAS.418.2584G,
       author = {{Ghosh}, Abhik and {Bharadwaj}, Somnath and {Ali}, Sk. Saiyad and {Chengalur}, Jayaram N.},
        title = "{Improved foreground removal in GMRT 610 MHz observations towards redshifted 21-cm tomography}",
      journal = {MNRAS},
         year = 2011,
        month = dec,
       volume = {418},
       number = {4},
        pages = {2584-2589},
          doi = {10.1111/j.1365-2966.2011.19649.x},
archivePrefix = {arXiv},
       eprint = {1108.3707},
 primaryClass = {astro-ph.CO},
       adsurl = {https://ui.adsabs.harvard.edu/abs/2011MNRAS.418.2584G}
}

@ARTICLE{poreion0,
   author = {{Wyithe}, J. Stuart B. and {Loeb}, Abraham},
        title = "{The 21-cm power spectrum after reionization}",
      journal = {Monthly Notices of the Royal Astronomical Society},
         year = 2009,
        month = aug,
       volume = {397},
       number = {4},
        pages = {1926-1934},
          doi = {10.1111/j.1365-2966.2009.15019.x},
archivePrefix = {arXiv},
       eprint = {0808.2323},
 primaryClass = {astro-ph},
       adsurl = {https://ui.adsabs.harvard.edu/abs/2009MNRAS.397.1926W}
}

@ARTICLE{poreion1,
   author = {{Bharadwaj}, S. and {Sethi}, S.~K.},
    title = "{HI Fluctuations at Large Redshifts: I--Visibility correlation}",
  journal = {Journal of Astrophysics and Astronomy},
   eprint = {arXiv:astro-ph/0203269},
     year = 2001,
    month = dec,
   volume = 22,
    pages = {293-307},
         adsurl = {http://adsabs.harvard.edu/abs/2001JApA...22..293B}
}

@ARTICLE{poreion2,
   author = {{Bharadwaj}, S. and {Nath}, B.~B. and {Sethi}, S.~K.},
    title = "{Using HI to Probe Large Scale Structures at z = 3}",
  journal = {Journal of Astrophysics and Astronomy},
   eprint = {arXiv:astro-ph/0003200},
     year = 2001,
    month = mar,
   volume = 22,
    pages = {21-32},
        adsurl = {http://adsabs.harvard.edu/abs/2001JApA...22...21B}
}

@ARTICLE{poreion3,
   author = {{Wyithe}, S. and {Loeb}, A.},
    title = "{Fluctuations in 21cm Emission After Reionization}",
  journal = {ArXiv e-prints},
archivePrefix = "arXiv",
   eprint = {0708.3392},
     year = 2007,
    month = aug,
   adsurl = {http://adsabs.harvard.edu/abs/2007arXiv0708.3392W}
}

@ARTICLE{poreion4,
   author = {{Loeb}, A. and {Wyithe}, J.~S.~B.},
    title = "{Possibility of Precise Measurement of the Cosmological Power Spectrum with a Dedicated Survey of 21cm Emission after Reionization}",
  journal = {Physical Review Letters},
archivePrefix = "arXiv",
   eprint = {0801.1677},
     year = 2008,
    month = apr,
   volume = 100,
   number = 16,
    pages = {161301-161310},
        adsurl = {http://adsabs.harvard.edu/abs/2008PhRvL.100p1301L}
}

@ARTICLE{poreion5,
   author = {{Wyithe}, S. and {Loeb}, A.},
    title = "{The 21cm Power Spectrum After Reionization}",
  journal = {ArXiv e-prints},
archivePrefix = "arXiv",
   eprint = {0808.2323},
     year = 2008,
    month = aug,
   adsurl = {http://adsabs.harvard.edu/abs/2008arXiv0808.2323W}
}

@ARTICLE{poreion6,
   author = {{Visbal}, E. and {Loeb}, A. and {Wyithe}, S.},
    title = "{Cosmological constraints from 21cm surveys after reionization}",
  journal = {Journal of Cosmology and Astro-Particle Physics},
archivePrefix = "arXiv",
   eprint = {0812.0419},
     year = 2009,
    month = oct,
   volume = 10,
    pages = {30-43},
        adsurl = {http://adsabs.harvard.edu/abs/2009JCAP...10..030V}
}

@ARTICLE{poreion7,
   author = {{Bharadwaj}, S. and {Pandey}, S.~K.},
    title = "{HI Fluctuations at Large Redshifts: II - the Signal Expected for the GMRT}",
  journal = {Journal of Astrophysics and Astronomy},
   eprint = {arXiv:astro-ph/0307303},
     year = 2003,
    month = mar,
   volume = 24,
    pages = {23-27},
   adsurl = {http://adsabs.harvard.edu/abs/2003JApA...24...23B}
}

@ARTICLE{poreion8,
   author = {{Bharadwaj}, S. and {Srikant}, P.~S.},
    title = "{HI Fluctuations at Large Redshifts: III - Simulating the Signal Expected at GMRT}",
  journal = {Journal of Astrophysics and Astronomy},
   eprint = {arXiv:astro-ph/0402262},
     year = 2004,
    month = mar,
   volume = 25,
    pages = {67-71},
      doi = {10.1007/BF02702289},
   adsurl = {http://adsabs.harvard.edu/abs/2004JApA...25...67B}
}

@ARTICLE{poreion9,
   author = {{Subramanian}, K. and {Padmanabhan}, T.},
    title = "{Neutral Hydrogen at High Redshifts as a Probe of Structure Formation - Part One - Post-Cobe Analysis of CDM and HDM Models}",
  journal = {MNRAS},
     year = 1993,
    month = nov,
   volume = 265,
    pages = {101},
   adsurl = {http://adsabs.harvard.edu/abs/1993MNRAS.265..101S}
}

@ARTICLE{poreion10,
   author = {{Kumar}, A. and {Padmanabhan}, T. and {Subramanian}, K.},
    title = "{Neutral hydrogen at high redshifts as a probe of structure formation - II. Line profile of a protocluster}",
  journal = {MNRAS},
     year = 1995,
    month = feb,
   volume = 272,
    pages = {544-550},
   adsurl = {http://adsabs.harvard.edu/abs/1995MNRAS.272..544K}
}

@ARTICLE{poreion11,
   author = {{Bagla}, J.~S. and {Nath}, B. and {Padmanabhan}, T.},
    title = "{Neutral hydrogen at high redshifts as a probe of structure formation - III. Radio maps from N-body simulations}",
  journal = {MNRAS},
   eprint = {arXiv:astro-ph/9610267},
     year = 1997,
    month = aug,
   volume = 289,
    pages = {671-680},
   adsurl = {http://adsabs.harvard.edu/abs/1997MNRAS.289..671B}
}

@article{Blackadder2014,
  title = {Dark matter with two- and many-body decays and supernovae type Ia},
  author = {Blackadder, Gordon and Koushiappas, Savvas M.},
  journal = {Phys. Rev. D},
  volume = {90},
  issue = {10},
  pages = {103527},
  numpages = {10},
  year = {2014},
  month = {Nov},
  publisher = {American Physical Society},
  doi = {10.1103/PhysRevD.90.103527},
  url = {https://link.aps.org/doi/10.1103/PhysRevD.90.103527}
}

@article{vattis2019late,
  title={Late universe decaying dark matter can relieve the h\_0 tension},
  author={Vattis, Kyriakos and Koushiappas, Savvas M and Loeb, Abraham},
  journal={arXiv preprint arXiv:1903.06220},
  year={2019}
}

@book{amendola_tsujikawa_2010, place={Cambridge}, title={Dark Energy: Theory and Observations}, DOI={10.1017/CBO9780511750823}, publisher={Cambridge University Press}, author={Amendola, Luca and Tsujikawa, Shinji}, year={2010}}

@misc{durrer2008dark,
    title={Dark Energy and Modified Gravity},
    author={Ruth Durrer and Roy Maartens},
    year={2008},
    eprint={0811.4132},
    archivePrefix={arXiv},
    primaryClass={astro-ph}
}

@article{CHEVALLIER_2001,
   title={ACCELERATING UNIVERSES WITH SCALING DARK MATTER},
   volume={10},
   ISSN={1793-6594},
   url={http://dx.doi.org/10.1142/S0218271801000822},
   DOI={10.1142/s0218271801000822},
   number={02},
   journal={International Journal of Modern Physics D},
   publisher={World Scientific Pub Co Pte Lt},
   author={CHEVALLIER, MICHEL and POLARSKI, DAVID},
   year={2001},
   month={Apr},
   pages={213–223}
}

@article{Zhao_2017,
   title={Dynamical dark energy in light of the latest observations},
   volume={1},
   ISSN={2397-3366},
   url={http://dx.doi.org/10.1038/s41550-017-0216-z},
   DOI={10.1038/s41550-017-0216-z},
   number={9},
   journal={Nature Astronomy},
   publisher={Springer Science and Business Media LLC},
   author={Zhao, Gong-Bo and Raveri, Marco and Pogosian, Levon and Wang, Yuting and Crittenden, Robert G. and Handley, Will J. and Percival, Will J. and Beutler, Florian and Brinkmann, Jonathan and Chuang, Chia-Hsun and et al.},
   year={2017},
   month={Aug},
   pages={627–632}
}

@article{Ratra-Peebles_1988,
  title = {Cosmological consequences of a rolling homogeneous scalar field},
  author = {Ratra, Bharat and Peebles, P. J. E.},
  journal = {Phys. Rev. D},
  volume = {37},
  issue = {12},
  pages = {3406--3427},
  numpages = {0},
  year = {1988},
  month = {Jun},
  publisher = {American Physical Society},
  doi = {10.1103/PhysRevD.37.3406},
  url = {https://link.aps.org/doi/10.1103/PhysRevD.37.3406}
}

@article{Steinhardt_1998,
  title = {Cosmological Imprint of an Energy Component with General Equation of State},
  author = {Caldwell, R. R. and Dave, Rahul and Steinhardt, Paul J.},
  journal = {Phys. Rev. Lett.},
  volume = {80},
  issue = {8},
  pages = {1582--1585},
  numpages = {0},
  year = {1998},
  month = {Feb},
  publisher = {American Physical Society},
  doi = {10.1103/PhysRevLett.80.1582},
  url = {https://link.aps.org/doi/10.1103/PhysRevLett.80.1582}
}

@misc{scranton2003physical,
    title={Physical Evidence for Dark Energy},
    author={R. Scranton and A. J. Connolly and R. C. Nichol and A. Stebbins and I. Szapudi and D. J. Eisenstein and N. Afshordi and T. Budavari and I. Csabai and J. A. Frieman and J. E. Gunn and D. Johnston and Y. Loh and R. H. Lupton and C. J. Miller and E. S. Sheldon and R. S. Sheth and A. S. Szalay and M. Tegmark and Y. Xu},
    year={2003},
    eprint={astro-ph/0307335},
    archivePrefix={arXiv},
    primaryClass={astro-ph}
       
}

@ARTICLE{bharad04,
   author = {{Bharadwaj}, S. and {Ali}, S.~S.},
    title = "{The cosmic microwave background radiation fluctuations from HI perturbations prior to reionization}",
  journal = {MNRAS},
   eprint = {arXiv:astro-ph/0401206},
     year = 2004,
    month = jul,
   volume = 352,
    pages = {142-146},
        adsurl = {http://adsabs.harvard.edu/abs/2004MNRAS.352..142B}
}

@ARTICLE{param1,
   author = {{Wyithe}, S. and {Loeb}, A. and {Geil}, P.},
    title = "{Baryonic Acoustic Oscillations in 21cm Emission: A Probe of Dark Energy out to High Redshifts}",
  journal = {ArXiv e-prints},
archivePrefix = "arXiv",
   eprint = {0709.2955},
     year = 2007,
    month = sep,
   adsurl = {http://adsabs.harvard.edu/abs/2007arXiv0709.2955W}
}

@ARTICLE{param2,
   author = {{Chang}, {T.-C.} and {Pen}, {U.-L.} and {Peterson}, J.~B. and 
	{McDonald}, P.},
    title = "{Baryon Acoustic Oscillation Intensity Mapping of Dark Energy}",
  journal = {Physical Review Letters},
archivePrefix = "arXiv",
   eprint = {0709.3672},
     year = 2008,
    month = mar,
   volume = 100,
   number = 9,
    pages = {091303-091310},
        adsurl = {http://adsabs.harvard.edu/abs/2008PhRvL.100i1303C}
}

@ARTICLE{param3,
   author = {{Bharadwaj}, S. and {Sethi}, S.~K. and {Saini}, T.~D.},
    title = "{Estimation of cosmological parameters from neutral hydrogen observations of the post-reionization epoch}",
  journal = {Physical Rev D},
archivePrefix = "arXiv",
   eprint = {0809.0363},
     year = 2009,
    month = apr,
   volume = 79,
   number = 8,
    pages = {083538-083542},
        adsurl = {http://adsabs.harvard.edu/abs/2009PhRvD..79h3538B}
}

@ARTICLE{param4,
   author = {{Mao}, Y. and {Tegmark}, M. and {McQuinn}, M. and {Zaldarriaga}, M. and 
	{Zahn}, O.},
    title = "{How accurately can 21cm tomography constrain cosmology?}",
  journal = {Physical Rev D},
archivePrefix = "arXiv",
   eprint = {0802.1710},
     year = 2008,
    month = jul,
   volume = 78,
   number = 2,
    pages = {023529-023532},
        adsurl = {http://adsabs.harvard.edu/abs/2008PhRvD..78b3529M}
}

@ARTICLE{xhibar,
   author = {{Lanzetta}, Kenneth M. and {Wolfe}, Arthur M. and {Turnshek}, David A.},
        title = "{The IUE Survey for Damped Lyman- alpha and Lyman-Limit Absorption Systems: Evolution of the Gaseous Content of the Universe}",
      journal = {The Astrophysical Journal },
         year = 1995,
        month = feb,
       volume = {440},
        pages = {435},
          doi = {10.1086/175286},
       adsurl = {https://ui.adsabs.harvard.edu/abs/1995ApJ...440..435L}
}

@ARTICLE{xhibar1,
   author = {{Storrie-Lombardi}, L.~J. and {McMahon}, R.~G. and {Irwin}, M.~J.
	},
    title = "{Evolution of neutral gas at high redshift: implications for the epoch of galaxy formation}",
  journal = {MNRAS},
   eprint = {arXiv:astro-ph/9608147},
     year = 1996,
    month = dec,
   volume = 283,
    pages = {L79-L83},
   adsurl = {http://adsabs.harvard.edu/abs/1996MNRAS.283L..79S}
}

@ARTICLE{xhibar2,
   author = {{Peroux}, C. and {McMahon}, R.~G. and {Storrie-Lombardi}, L.~J. and 
	{Irwin}, M.~J.},
    title = "{The evolution of OmegaHI and the epoch of formation of damped Lyman-alpha absorbers}",
  journal = {MNRAS},
   eprint = {arXiv:astro-ph/0107045},
     year = 2003,
    month = dec,
   volume = 346,
    pages = {1103-1115},
        adsurl = {http://adsabs.harvard.edu/abs/2003MNRAS.346.1103P}
}

@ARTICLE{proch05,
   author = {{Prochaska}, J.~X. and {Herbert-Fort}, S. and {Wolfe}, A.~M.
	},
    title = "{The SDSS Damped Ly{$\alpha$} Survey: Data Release 3}",
  journal = {ApJ},
   eprint = {arXiv:astro-ph/0508361},
     year = 2005,
    month = dec,
   volume = 635,
    pages = {123-142},
       adsurl = {http://adsabs.harvard.edu/abs/2005ApJ...635..123P}
}

@article{Bull_2015,
   title={LATE-TIME COSMOLOGY WITH 21 cm INTENSITY MAPPING EXPERIMENTS},
   volume={803},
   ISSN={1538-4357},
   url={http://dx.doi.org/10.1088/0004-637X/803/1/21},
   DOI={10.1088/0004-637x/803/1/21},
   number={1},
   journal={The Astrophysical Journal},
   publisher={IOP Publishing},
   author={Bull, Philip and Ferreira, Pedro G. and Patel, Prina and Santos, Mário G.},
   year={2015},
   month={Apr},
   pages={21}
    
  
}

@article{bagla20,
   title={HI as a probe of the large-scale structure in the post-reionization universe},
   volume={407},
   ISSN={0035-8711},
   url={http://dx.doi.org/10.1111/j.1365-2966.2010.16933.x},
   DOI={10.1111/j.1365-2966.2010.16933.x},
   number={1},
   journal={Monthly Notices of the Royal Astronomical Society},
   publisher={Oxford University Press (OUP)},
   author={Bagla, J. S. and Khandai, Nishikanta and Datta, Kanan K.},
   year={2010},
   month={Jun},
   pages={567–580}
}

@article{Guha_Sarkar_2012,
   title={Constraining large-scale HI bias using redshifted 21-cm signal from the post-reionization epoch},
   volume={421},
   ISSN={0035-8711},
   url={http://dx.doi.org/10.1111/j.1365-2966.2012.20582.x},
   DOI={10.1111/j.1365-2966.2012.20582.x},
   number={4},
   journal={Monthly Notices of the Royal Astronomical Society},
   publisher={Oxford University Press (OUP)},
   author={Guha Sarkar, Tapomoy and Mitra, Sourav and Majumdar, Suman and Choudhury, Tirthankar Roy},
   year={2012},
   month={Mar},
   pages={3570–3578}
}

@article{Sarkar_2016,
   title={Modelling the post-reionization neutral hydrogen HI bias},
   volume={460},
   ISSN={1365-2966},
   url={http://dx.doi.org/10.1093/mnras/stw1111},
   DOI={10.1093/mnras/stw1111},
   number={4},
   journal={Monthly Notices of the Royal Astronomical Society},
   publisher={Oxford University Press (OUP)},
   author={Sarkar, Debanjan and Bharadwaj, Somnath and Anathpindika, S.},
   year={2016},
   month={May},
   pages={4310–4319}
}

@article{Mar_n_2010,
   title={MODELING THE LARGE-SCALE BIAS OF NEUTRAL HYDROGEN},
   volume={718},
   ISSN={1538-4357},
   url={http://dx.doi.org/10.1088/0004-637X/718/2/972},
   DOI={10.1088/0004-637x/718/2/972},
   number={2},
   journal={The Astrophysical Journal},
   publisher={IOP Publishing},
   author={Marín, Felipe A. and Gnedin, Nickolay Y. and Seo, Hee-Jong and Vallinotto, Alberto},
   year={2010},
   month={Jul},
   pages={972–980}
}

@ARTICLE{Skordis_Coupled_DE_2013,
       author = {{Pourtsidou}, A. and {Skordis}, C. and {Copeland}, E.~J.},
        title = "{Models of dark matter coupled to dark energy}",
      journal = {Physical Review D},
         year = 2013,
        month = oct,
       volume = {88},
       number = {8},
          eid = {083505},
        pages = {083505},
          doi = {10.1103/PhysRevD.88.083505},
       adsurl = {https://ui.adsabs.harvard.edu/abs/2013PhRvD..88h3505P}
}

@article{masui2013measurement,
  author = {{Masui}, K.~W. and {Switzer}, E.~R. and {Banavar}, N. and {Bandura}, K. and {Blake}, C. and {Calin}, L. -M. and {Chang}, T. -C. and {Chen}, X. and {Li}, Y. -C. and {Liao}, Y. -W. and {Natarajan}, A. and {Pen}, U. -L. and {Peterson}, J.~B. and {Shaw}, J.~R. and {Voytek}, T.~C.},
        title = "{Measurement of 21 cm Brightness Fluctuations at z \raisebox{-0.5ex}\textasciitilde 0.8 in Cross-correlation}",
      journal = {The Astrophysical Journal Letters},
         year = 2013,
        month = jan,
       volume = {763},
       number = {1},
          eid = {L20},
        pages = {L20},
          doi = {10.1088/2041-8205/763/1/L20},
archivePrefix = {arXiv},
       eprint = {1208.0331},
 primaryClass = {astro-ph.CO},
       adsurl = {https://ui.adsabs.harvard.edu/abs/2013ApJ...763L..20M}
}

@article{di2002radio,
  author = "Di Matteo, Tiziana and Perna, Rosalba and Abel, Tom and Rees, Martin J.",
    title = "{Radio foregrounds for the 21cm tomography of the neutral intergalactic medium at high redshifts}",
    eprint = "astro-ph/0109241",
    archivePrefix = "arXiv",
    doi = "10.1086/324293",
    journal = "Astrophys. J.",
    volume = "564",
    pages = "576--580",
    year = "2002"
}

@article{shaver1999can,
   author = {{Shaver}, P.~A. and {Windhorst}, R.~A. and {Madau}, P. and {de Bruyn}, A.~G.},
        title = "{Can the reionization epoch be detected as a global signature in the cosmic background?}",
      journal = {Astronomy and Astrophysics},
         year = 1999,
        month = may,
       volume = {345},
        pages = {380-390},
          doi = {10.48550/arXiv.astro-ph/9901320},
archivePrefix = {arXiv},
       eprint = {astro-ph/9901320},
 primaryClass = {astro-ph},
       adsurl = {https://ui.adsabs.harvard.edu/abs/1999A&A...345..380S}
}

@article{pober2013opening,
   author = {{Pober}, Jonathan C. and {Parsons}, Aaron R. and {Aguirre}, James E. and {Ali}, Zaki and {Bradley}, Richard F. and {Carilli}, Chris L. and {DeBoer}, Dave and {Dexter}, Matthew and {Gugliucci}, Nicole E. and {Jacobs}, Daniel C. and {Klima}, Patricia J. and {MacMahon}, Dave and {Manley}, Jason and {Moore}, David F. and {Stefan}, Irina I. and {Walbrugh}, William P.},
        title = "{Opening the 21 cm Epoch of Reionization Window: Measurements of Foreground Isolation with PAPER}",
      journal = {Astrophysical Journal Letters},
         year = 2013,
        month = may,
       volume = {768},
       number = {2},
          eid = {L36},
        pages = {L36},
          doi = {10.1088/2041-8205/768/2/L36},
archivePrefix = {arXiv},
       eprint = {1301.7099},
 primaryClass = {astro-ph.CO},
       adsurl = {https://ui.adsabs.harvard.edu/abs/2013ApJ...768L..36P}
}

@article{pober2014next,
  author = {{Pober}, Jonathan C. and {Liu}, Adrian and {Dillon}, Joshua S. and {Aguirre}, James E. and {Bowman}, Judd D. and {Bradley}, Richard F. and {Carilli}, Chris L. and {DeBoer}, David R. and {Hewitt}, Jacqueline N. and {Jacobs}, Daniel C. and {McQuinn}, Matthew and {Morales}, Miguel F. and {Parsons}, Aaron R. and {Tegmark}, Max and {Werthimer}, Dan J.},
        title = "{What Next-generation 21 cm Power Spectrum Measurements can Teach us About the Epoch of Reionization}",
      journal = {The Astrophysical Journal},
         year = 2014,
        month = feb,
       volume = {782},
       number = {2},
          eid = {66},
        pages = {66},
          doi = {10.1088/0004-637X/782/2/66},
archivePrefix = {arXiv},
       eprint = {1310.7031},
 primaryClass = {astro-ph.CO},
       adsurl = {https://ui.adsabs.harvard.edu/abs/2014ApJ...782...66P}
}

@article{liu2014epoch,
  author = {{Liu}, Adrian and {Parsons}, Aaron R. and {Trott}, Cathryn M.},
        title = "{Epoch of reionization window. I. Mathematical formalism}",
      journal = {Physical Review D},
         year = 2014,
        month = jul,
       volume = {90},
       number = {2},
          eid = {023018},
        pages = {023018},
          doi = {10.1103/PhysRevD.90.023018},
archivePrefix = {arXiv},
       eprint = {1404.2596},
 primaryClass = {astro-ph.CO},
       adsurl = {https://ui.adsabs.harvard.edu/abs/2014PhRvD..90b3018L}
}

@article{dillon2015empirical,
    author = {{Dillon}, Joshua S. and et. al.},
        title = "{Empirical covariance modeling for 21 cm power spectrum estimation: A method demonstration and new limits from early Murchison Widefield Array 128-tile data}",
      journal = {Physical Review D},
         year = 2015,
        month = jun,
       volume = {91},
       number = {12},
          eid = {123011},
        pages = {123011},
          doi = {10.1103/PhysRevD.91.123011},
archivePrefix = {arXiv},
       eprint = {1506.01026},
 primaryClass = {astro-ph.CO},
       adsurl = {https://ui.adsabs.harvard.edu/abs/2015PhRvD..91l3011D}
}

@article{pal2021demonstrating,
   author = {{Pal}, Srijita and {Bharadwaj}, Somnath and {Ghosh}, Abhik and {Choudhuri}, Samir},
        title = "{Demonstrating the Tapered Gridded Estimator (TGE) for the cosmological H I 21-cm power spectrum using 150-MHz GMRT observations}",
      journal = {Monthly Notices of the Royal Astronomical Society},
         year = 2021,
        month = mar,
       volume = {501},
       number = {3},
        pages = {3378-3391},
          doi = {10.1093/mnras/staa3831},
archivePrefix = {arXiv},
       eprint = {2012.04998},
 primaryClass = {astro-ph.CO},
       adsurl = {https://ui.adsabs.harvard.edu/abs/2021MNRAS.501.3378P}
}

@article{mcquinn2006cosmological,
   author = {{McQuinn}, Matthew and {Zahn}, Oliver and {Zaldarriaga}, Matias and {Hernquist}, Lars and {Furlanetto}, Steven R.},
        title = "{Cosmological Parameter Estimation Using 21 cm Radiation from the Epoch of Reionization}",
      journal = {The Astrophysical Journal},
         year = 2006,
        month = dec,
       volume = {653},
       number = {2},
        pages = {815-834},
          doi = {10.1086/505167},
archivePrefix = {arXiv},
       eprint = {astro-ph/0512263},
 primaryClass = {astro-ph},
       adsurl = {https://ui.adsabs.harvard.edu/abs/2006ApJ...653..815M}
}

@ARTICLE{Potter,
       author = {{Potter}, Douglas and {Stadel}, Joachim and {Teyssier}, Romain},
        title = "{PKDGRAV3: beyond trillion particle cosmological simulations for the next era of galaxy surveys}",
      journal = {Computational Astrophysics and Cosmology},
         year = 2017,
        month = may,
       volume = {4},
       number = {1},
          eid = {2},
        pages = {2},
          doi = {10.1186/s40668-017-0021-1},
archivePrefix = {arXiv},
       eprint = {1609.08621},
 primaryClass = {astro-ph.IM},
       adsurl = {https://ui.adsabs.harvard.edu/abs/2017ComAC...4....2P}
}

@ARTICLE{Peters,
       author = {{Bucko}, Jozef and {Giri}, Sambit K. and {Peters}, Fabian Hervas and {Schneider}, Aurel},
        title = "{Probing the two-body decaying dark matter scenario with weak lensing and the cosmic microwave background}",
      journal = {Astronomy and Astrophysics},
         year = 2024,
        month = mar,
       volume = {683},
          eid = {A152},
        pages = {A152},
          doi = {10.1051/0004-6361/202347844},
archivePrefix = {arXiv},
       eprint = {2307.03222},
 primaryClass = {astro-ph.CO},
       adsurl = {https://ui.adsabs.harvard.edu/abs/2024A&A...683A.152B}
}

@article{jackson1972critique,
  title={A critique of Rees's theory of primordial gravitational radiation},
  author={Jackson, JC},
  journal={Monthly Notices of the Royal Astronomical Society},
  volume={156},
  number={1},
  pages={1P--5P},
  year={1972},
  publisher={Oxford University Press},
  doi = {10.1093/mnras/156.1.1P}
}

@ARTICLE{AP1979,
       author = {{Alcock}, C. and {Paczynski}, B.},
        title = "{An evolution free test for non-zero cosmological constant}",
      journal = {Nature},
         year = 1979,
        month = oct,
       volume = {281},
        pages = {358},
          doi = {10.1038/281358a0},
       adsurl = {https://ui.adsabs.harvard.edu/abs/1979Natur.281..358A}
}

@article{lopez2014alcock,
  title={ALCOCK--PACZYNSKI COSMOLOGICAL TEST},
  author={Lopez-Corredoira, Martin},
  journal={The Astrophysical Journal},
  volume={781},
  number={2},
  pages={96},
  year={2014},
  publisher={IOP Publishing},
  doi = {10.1088/0004-637X/781/2/96}
}

@ARTICLE{ddmdata,
       author = {{Fu{\ss}}, Lea and {Garny}, Mathias},
        title = "{Decaying Dark Matter and Lyman-{\ensuremath{\alpha}} forest constraints}",
      journal = {Journal of Cosmology and Astroparticle Physics},
         year = 2023,
        month = oct,
       volume = {2023},
       number = {10},
          eid = {020},
        pages = {020},
          doi = {10.1088/1475-7516/2023/10/020},
archivePrefix = {arXiv},
       eprint = {2210.06117},
 primaryClass = {astro-ph.CO},
       adsurl = {https://ui.adsabs.harvard.edu/abs/2023JCAP...10..020F}
}

@ARTICLE{shotnoise,
       author = {{Villaescusa-Navarro}, Francisco and {Genel}, Shy and {Castorina}, Emanuele and {Obuljen}, Andrej and {Spergel}, David N. and {Hernquist}, Lars and {Nelson}, Dylan and {Carucci}, Isabella P. and {Pillepich}, Annalisa and {Marinacci}, Federico and {Diemer}, Benedikt and {Vogelsberger}, Mark and {Weinberger}, Rainer and {Pakmor}, R{\"u}diger},
        title = "{Ingredients for 21 cm Intensity Mapping}",
      journal = {\apj},
         year = 2018,
        month = oct,
       volume = {866},
       number = {2},
          eid = {135},
        pages = {135},
          doi = {10.3847/1538-4357/aadba0},
archivePrefix = {arXiv},
       eprint = {1804.09180},
 primaryClass = {astro-ph.CO},
       adsurl = {https://ui.adsabs.harvard.edu/abs/2018ApJ...866..135V}
}

@ARTICLE{pankaj1,
       author = {{Chavan}, Pankaj and {Guha Sarkar}, Tapomoy and {Dash}, Chandrachud B.~V. and {Sen}, Anjan A.},
        title = "{A semi-cosmographic approach to study cosmological evolution in phase space}",
      journal = {\jcap},
         year = 2025,
        month = jul,
       volume = {2025},
       number = {7},
          eid = {029},
        pages = {029},
          doi = {10.1088/1475-7516/2025/07/029},
archivePrefix = {arXiv},
       eprint = {2503.03288},
 primaryClass = {astro-ph.CO},
       adsurl = {https://ui.adsabs.harvard.edu/abs/2025JCAP...07..029C}
}

@ARTICLE{pankaj2,
       author = {{Chavan}, Pankaj and {Guha Sarkar}, Tapomoy and {Sen}, Anjan A.},
        title = "{The dynamics of background cosmological evolution and structure formation in phase space: A semi-cosmographic reconstruction}",
      journal = {Physics of the Dark Universe},
         year = 2025,
        month = dec,
       volume = {50},
          eid = {102173},
        pages = {102173},
          doi = {10.1016/j.dark.2025.102173},
archivePrefix = {arXiv},
       eprint = {2506.14275},
 primaryClass = {astro-ph.CO},
       adsurl = {https://ui.adsabs.harvard.edu/abs/2025PDU....5002173C}
}

@article{Saini_2000,
   title={Reconstructing the Cosmic Equation of State from Supernova Distances},
   volume={85},
   ISSN={1079-7114},
   url={http://dx.doi.org/10.1103/PhysRevLett.85.1162},
   DOI={10.1103/physrevlett.85.1162},
   number={6},
   journal={Physical Review Letters},
   publisher={American Physical Society (APS)},
   author={Saini, Tarun Deep and Raychaudhury, Somak and Sahni, Varun and Starobinsky, A. A.},
   year={2000},
   month=aug, pages={1162–1165} }

@ARTICLE{poreion12,
	author = {{Padmanabhan}, H and T. Roy Choudhury and Alexandre Refregier},
  	title = "{Theoretical and observational constraints on the Hi intensity power spectrum}",
  	journal = {Monthly Notices of the Royal Astronomical Society},
  	year = 2015,
	month = jan, 
  	volume = {447},
  	number = {4},
  	pages = {3745--3755},
doi = {10.1093/mnras/stu2702}
}

@article{scherrer2008thawing,
  title = {Thawing quintessence with a nearly flat potential},
  author = {Scherrer, Robert J. and Sen, A. A.},
  journal = {Phys. Rev. D},
  volume = {77},
  issue = {8},
  pages = {083515},
  numpages = {8},
  year = {2008},
  month = {Apr},
  publisher = {American Physical Society},
  doi = {10.1103/PhysRevD.77.083515},
  url = {https://link.aps.org/doi/10.1103/PhysRevD.77.083515}
}

@article{amendola2000coupled,
  title={Coupled quintessence},
  author={Amendola, Luca},
  journal={Physical Review D},
  volume={62},
  number={4},
  pages={043511},
  year={2000},
  publisher={APS}
}

@article{PhysRevLett.82.896,
  title = {Quintessence, Cosmic Coincidence, and the Cosmological Constant},
  author = {Zlatev, Ivaylo and Wang, Limin and Steinhardt, Paul J.},
  journal = {Phys. Rev. Lett.},
  volume = {82},
  issue = {5},
  pages = {896--899},
  numpages = {0},
  year = {1999},
  month = {Feb},
  publisher = {American Physical Society},
  doi = {10.1103/PhysRevLett.82.896},
  url = {https://link.aps.org/doi/10.1103/PhysRevLett.82.896}
}

@book{Weinberg_1972_cosmography,
    author = "Weinberg, Steven",
    title = "{Gravitation and Cosmology}: {Principles and Applications of the General Theory of Relativity}",
    isbn = "978-0-471-92567-5, 978-0-471-92567-5",
    publisher = "John Wiley and Sons",
    address = "New York",
    year = "1972"
}

@article{Shafieloo_2012_GPR,
  title = {Gaussian process cosmography},
  author = {Shafieloo, Arman and Kim, Alex G. and Linder, Eric V.},
  journal = {Phys. Rev. D},
  volume = {85},
  issue = {12},
  pages = {123530},
  numpages = {9},
  year = {2012},
  month = {Jun},
  publisher = {American Physical Society},
  doi = {10.1103/PhysRevD.85.123530},
  url = {https://link.aps.org/doi/10.1103/PhysRevD.85.123530}
}

@Article{Velazquez_mukherjee_GPR_2024,
AUTHOR = {Velázquez, José de Jesús and Escamilla, Luis A. and Mukherjee, Purba and Vázquez, J. Alberto},
TITLE = {Non-Parametric Reconstruction of Cosmological Observables Using Gaussian Processes Regression},
JOURNAL = {Universe},
VOLUME = {10},
YEAR = {2024},
NUMBER = {12},
ARTICLE-NUMBER = {464},
URL = {https://www.mdpi.com/2218-1997/10/12/464},
ISSN = {2218-1997},
DOI = {10.3390/universe10120464}
}

@article{Jesus_2024_GPR,
    author = {Jesus, J F and Benndorf, D and Escobal, A A and Pereira, S H},
    title = {From Hubble to snap parameters: a Gaussian process reconstruction},
    journal = {Monthly Notices of the Royal Astronomical Society},
    volume = {528},
    number = {2},
    pages = {1573-1581},
    year = {2024},
    month = {01},
    issn = {0035-8711},
    doi = {10.1093/mnras/stae120},
    url = {https://doi.org/10.1093/mnras/stae120},
    eprint ={https://academic.oup.com/mnras/article-pdf/528/2/1573/56410686/stae120.pdf},
}

@ARTICLE{Dinda2024_GP_cosmography,
  title    = "Analytical Gaussian process cosmography: unveiling insights into matter-energy density parameter at present",
  author   = "Dinda, Bikash R",
  journal  = "The European Physical Journal C",
  volume   =  84,
  number   =  4,
  pages    = "402",
  month    =  apr,
  year     =  2024
}

@article{Holsclaw_2011_GPR,
  title = {Nonparametric reconstruction of the dark energy equation of state from diverse data sets},
  author = {Holsclaw, Tracy and Alam, Ujjaini and Sans\'o, Bruno and Lee, Herbie and Heitmann, Katrin and Habib, Salman and Higdon, David},
  journal = {Phys. Rev. D},
  volume = {84},
  issue = {8},
  pages = {083501},
  numpages = {18},
  year = {2011},
  month = {Oct},
  publisher = {American Physical Society},
  doi = {10.1103/PhysRevD.84.083501},
  url = {https://link.aps.org/doi/10.1103/PhysRevD.84.083501}
}

@article{Visser_2015_cosmography,
doi = {10.1088/0264-9381/32/13/135007},
url = {https://dx.doi.org/10.1088/0264-9381/32/13/135007},
year = {2015},
month = {jun},
publisher = {IOP Publishing},
volume = {32},
number = {13},
pages = {135007},
author = {Visser, Matt},
title = {Conformally Friedmann–Lemaître–Robertson–Walker cosmologies},
journal = {Classical and Quantum Gravity}
}

@article{Dunsby_Luongo_2016_cosmography,
author = {Dunsby, Peter K. S. and Luongo, Orlando},
title = {On the theory and applications of modern cosmography},
journal = {International Journal of Geometric Methods in Modern Physics},
volume = {13},
number = {03},
pages = {1630002},
year = {2016},
doi = {10.1142/S0219887816300026},
URL = {https://doi.org/10.1142/S0219887816300026}
}

@article{Capozziello_2019_cosmography,
author = {Capozziello, Salvatore and D’Agostino, Rocco and Luongo, Orlando},
title = {Extended gravity cosmography},
journal = {International Journal of Modern Physics D},
volume = {28},
number = {10},
pages = {1930016},
year = {2019},
doi = {10.1142/S0218271819300167},
URL = {https://doi.org/10.1142/S0218271819300167}
}

@article{Cattoen_2007_convergence,
doi = {10.1088/0264-9381/24/23/018},
url = {https://dx.doi.org/10.1088/0264-9381/24/23/018},
year = {2007},
month = {nov},
publisher = {},
volume = {24},
number = {23},
pages = {5985},
author = {Cattoën, Céline and Visser, Matt},
title = {The Hubble series: convergence properties and redshift variables},
journal = {Classical and Quantum Gravity}
}

@article{Pade_1892,
     author = {Pad\'e, H.},
     title = {Sur la repr\'esentation approch\'ee d'une fonction par des fractions rationnelles},
     journal = {Annales scientifiques de l'\'Ecole Normale Sup\'erieure},
     pages = {3--93},
     publisher = {Elsevier},
     volume = {3e s{\'e}rie, 9},
     year = {1892},
     doi = {10.24033/asens.378},
     language = {fr},
     url = {http://www.numdam.org/articles/10.24033/asens.378/}
}

@article{Aviles_2012_cosmography_y_variable,
  title = {Cosmography and constraints on the equation of state of the Universe in various parametrizations},
  author = {Aviles, Alejandro and Gruber, Christine and Luongo, Orlando and Quevedo, Hernando},
  journal = {Phys. Rev. D},
  volume = {86},
  issue = {12},
  pages = {123516},
  numpages = {23},
  year = {2012},
  month = {Dec},
  publisher = {American Physical Society},
  doi = {10.1103/PhysRevD.86.123516},
  url = {https://link.aps.org/doi/10.1103/PhysRevD.86.123516}
}

@article{Wei_2014_cosmography,
doi = {10.1088/1475-7516/2014/01/045},
url = {https://dx.doi.org/10.1088/1475-7516/2014/01/045},
year = {2014},
month = {jan},
publisher = {},
volume = {2014},
number = {01},
pages = {045},
author = {Wei, Hao and Yan, Xiao-Peng and Zhou, Ya-Nan},
title = {Cosmological applications of Padé approximant},
journal = {Journal of Cosmology and Astroparticle Physics}
}

@article{Aviles_2014_cosmography,
  title = {Precision cosmology with Pad\'e rational approximations: Theoretical predictions versus observational limits},
  author = {Aviles, Alejandro and Bravetti, Alessandro and Capozziello, Salvatore and Luongo, Orlando},
  journal = {Phys. Rev. D},
  volume = {90},
  issue = {4},
  pages = {043531},
  numpages = {25},
  year = {2014},
  month = {Aug},
  publisher = {American Physical Society},
  doi = {10.1103/PhysRevD.90.043531},
  url = {https://link.aps.org/doi/10.1103/PhysRevD.90.043531}
}

@article{Lobo_2020,
   title={Cosmographic analysis of redshift drift},
   volume={2020},
   ISSN={1475-7516},
   url={http://dx.doi.org/10.1088/1475-7516/2020/04/043},
   DOI={10.1088/1475-7516/2020/04/043},
   number={04},
   journal={Journal of Cosmology and Astroparticle Physics},
   publisher={IOP Publishing},
   author={Lobo, Francisco S.N. and Mimoso, José Pedro and Visser, Matt},
   year={2020},
   month=apr, pages={043–043} }

@ARTICLE{Zhou_2016_cosmography,
  title    = "New generalizations of cosmography inspired by the Pad{\'e} approximant",
  author   = "Zhou, Ya-Nan and Liu, De-Zi and Zou, Xiao-Bo and Wei, Hao",
  journal  = "The European Physical Journal C",
  volume   =  76,
  number   =  5,
  pages    = "281",
  month    =  may,
  year     =  2016
}

@article{Capozziello_Ruchika_Anjan_2019_cosmography,
    author = {Capozziello, Salvatore and Ruchika and Sen, Anjan A},
    title = {Model-independent constraints on dark energy evolution from low-redshift observations},
    journal = {Monthly Notices of the Royal Astronomical Society},
    volume = {484},
    number = {4},
    pages = {4484-4494},
    year = {2019},
    month = {01},
}

@article{Capozziello_2020_cosmography,
    author = {Capozziello, S and D’Agostino, R and Luongo, O},
    title = {High-redshift cosmography: auxiliary variables versus Padé polynomials},
    journal = {Monthly Notices of the Royal Astronomical Society},
    volume = {494},
    number = {2},
    pages = {2576-2590},
    year = {2020},
    month = {04}
}

@article{Pourojaghi2022,
  title = {Can high-redshift Hubble diagrams rule out the standard model of cosmology in the context of cosmography?},
  author = {Pourojaghi, S. and Zabihi, N. F. and Malekjani, M.},
  journal = {Phys. Rev. D},
  volume = {106},
  issue = {12},
  pages = {123523},
  numpages = {17},
  year = {2022},
  month = {Dec},
  publisher = {American Physical Society},
  doi = {10.1103/PhysRevD.106.123523},
  url = {https://link.aps.org/doi/10.1103/PhysRevD.106.123523}
}

@article{Petreca_2024,
title = {Beyond LCDM with f(z)CDM: Criticalities and solutions of Padé Cosmography},
journal = {Physics of the Dark Universe},
volume = {44},
pages = {101453},
year = {2024},
issn = {2212-6864},
doi = {https://doi.org/10.1016/j.dark.2024.101453},
url = {https://www.sciencedirect.com/science/article/pii/S2212686424000359},
author = {A. Turmina Petreca and M. Benetti and S. Capozziello}
}

@article{Busti_2015_cosmography,
  title = {Is cosmography a useful tool for testing cosmology?},
  author = {Busti, Vinicius C. and de la Cruz-Dombriz, \'Alvaro and Dunsby, Peter K. S. and S\'aez-G\'omez, Diego},
  journal = {Phys. Rev. D},
  volume = {92},
  issue = {12},
  pages = {123512},
  numpages = {8},
  year = {2015},
  month = {Dec},
  publisher = {American Physical Society},
  doi = {10.1103/PhysRevD.92.123512},
  url = {https://link.aps.org/doi/10.1103/PhysRevD.92.123512}
}

@ARTICLE{Visser2005-cosmography,
  title    = "Cosmography: Cosmology without the Einstein equations",
  author   = "Visser, Matt",
  journal  = "General Relativity and Gravitation",
  volume   =  37,
  number   =  9,
  pages    = "1541--1548",
  month    =  sep,
  year     =  2005
}

@article{Yang_Aritra_banerjee_2020_cosmography,
  title = {Cosmography and flat $\mathrm{\ensuremath{\Lambda}}\mathrm{CDM}$ tensions at high redshift},
  author = {Yang, Tao and Banerjee, Aritra and \'O Colg\'ain, Eoin},
  journal = {Phys. Rev. D},
  volume = {102},
  issue = {12},
  pages = {123532},
  numpages = {14},
  year = {2020},
  month = {Dec},
  publisher = {American Physical Society},
  doi = {10.1103/PhysRevD.102.123532},
  url = {https://link.aps.org/doi/10.1103/PhysRevD.102.123532}
}

@article{Aviles_2013_cosmography,
  title = {Updated constraints on $f(\mathcal{R})$ gravity from cosmography},
  author = {Aviles, Alejandro and Bravetti, Alessandro and Capozziello, Salvatore and Luongo, Orlando},
  journal = {Phys. Rev. D},
  volume = {87},
  issue = {4},
  pages = {044012},
  numpages = {11},
  year = {2013},
  month = {Feb},
  publisher = {American Physical Society},
  doi = {10.1103/PhysRevD.87.044012},
  url = {https://link.aps.org/doi/10.1103/PhysRevD.87.044012}
}

@article{Aviles_bravetti_cosmography_2013,
  title = {Cosmographic reconstruction of $f(\mathcal{T})$ cosmology},
  author = {Aviles, Alejandro and Bravetti, Alessandro and Capozziello, Salvatore and Luongo, Orlando},
  journal = {Phys. Rev. D},
  volume = {87},
  issue = {6},
  pages = {064025},
  numpages = {10},
  year = {2013},
  month = {Mar},
  publisher = {American Physical Society},
  doi = {10.1103/PhysRevD.87.064025},
  url = {https://link.aps.org/doi/10.1103/PhysRevD.87.064025}
}

@ARTICLE{Mehrabi_2018_cosmography,
  title    = {Dark energy reconstruction based on the Pade approximation an expansion around the LCDM},
  author   = {Mehrabi, Ahmad and Basilakos, Spyros},
  journal  = {The European Physical Journal C},
  volume   =  {78},
  number   =  {11},
  pages    = {889},
  month    =  {nov},
  year     =  {2018}
}

@article{Rezaei_2017_cosmography,
doi = {10.3847/1538-4357/aa7898},
url = {https://dx.doi.org/10.3847/1538-4357/aa7898},
year = {2017},
month = {jul},
publisher = {The American Astronomical Society},
volume = {843},
number = {1},
pages = {65},
author = {Rezaei, M. and Malekjani, M. and Basilakos, S. and Mehrabi, A. and Mota, D. F.},
title = {Constraints to Dark Energy Using PADE Parameterizations},
journal = {The Astrophysical Journal}
}

@ARTICLE{Liu_2021_cosmography,
  title  = {Bias of reconstructing the dark energy equation of state from the Pad{\'e} cosmography},
  author   = {Liu, Yang and Li, Zhengxiang and Yu, Hongwei and Wu, Puxun},
  journal  = {Astrophysics and Space Science},
  volume   =  {366},
  number   =  {11},
  pages    = {112},
  month    =  {nov},
  year     =  {2021}
}

@article{Capozziello_2018_cosmography,
doi = {10.1088/1475-7516/2018/05/008},
url = {https://dx.doi.org/10.1088/1475-7516/2018/05/008},
year = {2018},
month = {may},
publisher = {},
volume = {2018},
number = {05},
pages = {008},
author = {Capozziello, Salvatore and D'Agostino, Rocco and Luongo, Orlando},
title = {Rational approximations of f(R) cosmography  through Pad'e polynomials},
journal = {Journal of Cosmology and Astroparticle Physics}
}

@article{Benetti_2019_cosmography,
doi = {10.1088/1475-7516/2019/12/008},
url = {https://dx.doi.org/10.1088/1475-7516/2019/12/008},
year = {2019},
month = {dec},
publisher = {},
volume = {2019},
number = {12},
pages = {008},
author = {Benetti, Micol and Capozziello, Salvatore},
title = {Connecting early and late epochs by f(z)CDM cosmography},
journal = {Journal of Cosmology and Astroparticle Physics}
}

@article{Dinda_2025,
doi = {10.1088/1475-7516/2025/01/120},
url = {https://dx.doi.org/10.1088/1475-7516/2025/01/120},
year = {2025},
month = {jan},
publisher = {IOP Publishing},
volume = {2025},
number = {01},
pages = {120},
author = {Dinda, Bikash R. and Maartens, Roy},
title = {Model-agnostic assessment of dark energy after DESI DR1 BAO},
journal = {Journal of Cosmology and Astroparticle Physics}
}

@article{Purba_Mukherjee-Anjan_sen_GPR_2024,
  title = {Model-independent cosmological inference post DESI DR1 BAO measurements},
  author = {Mukherjee, Purba and Sen, Anjan Ananda},
  journal = {Phys. Rev. D},
  volume = {110},
  issue = {12},
  pages = {123502},
  numpages = {11},
  year = {2024},
  month = {Dec},
  publisher = {American Physical Society},
  doi = {10.1103/PhysRevD.110.123502},
  url = {https://link.aps.org/doi/10.1103/PhysRevD.110.123502}
}

@article{DESI_Colab_2025_DR2,
  author = {Abdul Karim, M. and others},
  collaboration = {DESI Collaboration},
  title = {DESI DR2 results. II. Measurements of baryon acoustic oscillations and cosmological constraints},
  journal = {Phys. Rev. D},
  volume = {112},
  pages = {083515},
  year = {2025},
  doi = {10.1103/tr6y-kpc6},
  url = {https://link.aps.org/doi/10.1103/tr6y-kpc6}
}

@ARTICLE{rubin,
       author = {{Rubin}, Vera C. and {Ford}, W. Kent Jr. and {Thonnard}, Norbert},
        title = "{Rotational Properties of 21 SC Galaxies with a Large Range of Luminosities and Radii}",
      journal = {The Astrophysical Journal},
         year = 1980,
       volume = {238},
        pages = {471-487},
          doi = {10.1086/158003},
       adsurl = {https://ui.adsabs.harvard.edu/abs/1980ApJ...238..471R}
}

@ARTICLE{bosma,
       author = {{Bosma}, A.},
        title = "{21-cm line studies of spiral galaxies. II. The distribution and kinematics of neutral hydrogen in spiral galaxies}",
      journal = {The Astronomical Journal},
         year = 1981,
       volume = {86},
        pages = {1825-1846},
          doi = {10.1086/113063},
       adsurl = {https://ui.adsabs.harvard.edu/abs/1981AJ.....86.1825B}
}

@ARTICLE{sofue,
       author = {{Sofue}, Yoshiaki and {Rubin}, Vera},
        title = "{Rotation Curves of Spiral Galaxies}",
      journal = {Annual Review of Astronomy and Astrophysics},
         year = 2001,
       volume = {39},
        pages = {137-174},
          doi = {10.1146/annurev.astro.39.1.137},
       adsurl = {https://ui.adsabs.harvard.edu/abs/2001ARA&A..39..137S}
}

@ARTICLE{lelli,
       author = {{Lelli}, Federico and {McGaugh}, Stacy S. and {Schombert}, James M.},
        title = "{SPARC: Mass Models for 175 Disk Galaxies with Spitzer Photometry and Accurate Rotation Curves}",
      journal = {The Astronomical Journal},
         year = 2016,
       volume = {152},
        pages = {157},
          doi = {10.3847/0004-6256/152/6/157},
       adsurl = {https://ui.adsabs.harvard.edu/abs/2016AJ....152..157L}
}

@ARTICLE{smoot,
  author = {{Smoot}, G.~F. and {Bennett}, C.~L. and {Kogut}, A. and et al.},
  title = "{Structure in the COBE Differential Microwave Radiometer First-Year Maps}",
  journal = {Astrophysical Journal Letters},
  year = 1992,
  volume = {396},
  pages = {L1-L5},
  doi = {10.1086/186504},
  adsurl = {https://ui.adsabs.harvard.edu/abs/1992ApJ...396L...1S}
}

@ARTICLE{dodleson,
  author = {{Hu}, Wayne and {Dodelson}, Scott},
  title = "{Cosmic Microwave Background Anisotropies}",
  journal = {Annual Review of Astronomy and Astrophysics},
  year = 2002,
  volume = {40},
  pages = {171-216},
  doi = {10.1146/annurev.astro.40.060401.093926},
  adsurl = {https://ui.adsabs.harvard.edu/abs/2002ARA&A..40..171H}
}

@BOOK{peebles,
  author = {{Peebles}, P.~J.~E.},
  title = "{The Large-Scale Structure of the Universe}",
  publisher = {Princeton University Press},
  year = 1980,
  adsurl = {https://ui.adsabs.harvard.edu/abs/1980lssu.book.....P}
}

@BOOK{peacock,
  author = {{Peacock}, J.~A.},
  title = "{Cosmological Physics}",
  publisher = {Cambridge University Press},
  year = 1999,
  adsurl = {https://ui.adsabs.harvard.edu/abs/1999coph.book.....P}
}

@ARTICLE{weinberg,
  author = {{Weinberg}, D.~H. and {Mortonson}, M.~J. and {Eisenstein}, D.~J. and et al.},
  title = "{Observational probes of cosmic acceleration}",
  journal = {Physics Reports},
  year = 2013,
  volume = {530},
  pages = {87-255},
  doi = {10.1016/j.physrep.2013.05.001},
  adsurl = {https://ui.adsabs.harvard.edu/abs/2013PhR...530...87W}
}

@ARTICLE{riess,
  author = {{Riess}, A.~G. and {Filippenko}, A.~V. and {Challis}, P. and et al.},
  title = "{Observational Evidence from Supernovae for an Accelerating Universe and a Cosmological Constant}",
  journal = {Astronomical Journal},
  year = 1998,
  volume = {116},
  pages = {1009-1038},
  doi = {10.1086/300499},
  adsurl = {https://ui.adsabs.harvard.edu/abs/1998AJ....116.1009R}
}

@ARTICLE{perlmutter,
  author = {{Perlmutter}, S. and {Aldering}, G. and {Goldhaber}, G. and et al.},
  title = "{Measurements of Omega and Lambda from 42 High-Redshift Supernovae}",
  journal = {Astrophysical Journal},
  year = 1999,
  volume = {517},
  pages = {565-586},
  doi = {10.1086/307221},
  adsurl = {https://ui.adsabs.harvard.edu/abs/1999ApJ...517..565P}
}

@ARTICLE{betoule,
  author = {{Betoule}, M. and {Kessler}, R. and {Guy}, J. and et al.},
  title = "{Improved cosmological constraints from a joint analysis of the SDSS-II and SNLS supernova samples}",
  journal = {Astronomy and Astrophysics},
  year = 2014,
  volume = {568},
  pages = {A22},
  doi = {10.1051/0004-6361/201423413},
  adsurl = {https://ui.adsabs.harvard.edu/abs/2014A&A...568A..22B}
}

@ARTICLE{sclonic,
  author = {{Scolnic}, D.~M. and {Jones}, D.~O. and {Rest}, A. and et al.},
  title = "{The Complete Light-curve Sample of Spectroscopically Confirmed Type Ia Supernovae from Pan-STARRS1}",
  journal = {Astrophysical Journal},
  year = 2018,
  volume = {859},
  pages = {101},
  doi = {10.3847/1538-4357/aab9bb},
  adsurl = {https://ui.adsabs.harvard.edu/abs/2018ApJ...859..101S}
}

@ARTICLE{eisenstein,
  author = {{Eisenstein}, D.~J. and {Zehavi}, I. and {Hogg}, D.~W. and et al.},
  title = "{Detection of the Baryon Acoustic Peak in the Large-Scale Correlation Function of SDSS Luminous Red Galaxies}",
  journal = {Astrophysical Journal},
  year = 2005,
  volume = {633},
  pages = {560-574},
  doi = {10.1086/466512},
  adsurl = {https://ui.adsabs.harvard.edu/abs/2005ApJ...633..560E}
}

@ARTICLE{anderson,
  author = {{Anderson}, L. and {Aubourg}, E. and {Bailey}, S. and et al.},
  title = "{The clustering of galaxies in the SDSS-III Baryon Oscillation Spectroscopic Survey}",
  journal = {Monthly Notices of the Royal Astronomical Society},
  year = 2014,
  volume = {441},
  pages = {24-62},
  doi = {10.1093/mnras/stu523},
  adsurl = {https://ui.adsabs.harvard.edu/abs/2014MNRAS.441...24A}
}

@ARTICLE{alam,
  author = {{Alam}, S. and {Aubourg}, E. and {Avila}, S. and et al.},
  title = "{Completed SDSS-IV extended Baryon Oscillation Spectroscopic Survey: Cosmological implications}",
  journal = {Physical Review D},
  year = 2021,
  volume = {103},
  pages = {083533},
  doi = {10.1103/PhysRevD.103.083533},
  adsurl = {https://ui.adsabs.harvard.edu/abs/2021PhRvD.103h3001A}
}

@ARTICLE{bartlemann,
  author = {{Bartelmann}, M. and {Schneider}, P.},
  title = "{Weak gravitational lensing}",
  journal = {Physics Reports},
  year = 2001,
  volume = {340},
  pages = {291-472},
  doi = {10.1016/S0370-1573(00)00082-X},
  adsurl = {https://ui.adsabs.harvard.edu/abs/2001PhR...340..291B}
}

@ARTICLE{kilbinger,
  author = {{Kilbinger}, M.},
  title = "{Cosmology with cosmic shear observations}",
  journal = {Reports on Progress in Physics},
  year = 2015,
  volume = {78},
  pages = {086901},
  doi = {10.1088/0034-4885/78/8/086901},
  adsurl = {https://ui.adsabs.harvard.edu/abs/2015RPPh...78h6901K}
}

@ARTICLE{abbott,
  author = {{Abbott}, T.~M.~C. and {Abdalla}, F.~B. and {Allam}, S. and et al.},
  title = "{Dark Energy Survey Year 1 Results: Cosmological Constraints from Galaxy Clustering and Weak Lensing}",
  journal = {Physical Review D},
  year = 2018,
  volume = {98},
  pages = {043526},
  doi = {10.1103/PhysRevD.98.043526},
  adsurl = {https://ui.adsabs.harvard.edu/abs/2018PhRvD..98d3516A}
}

@ARTICLE{refsdal,
  author = {{Refsdal}, S.},
  title = "{On the possibility of determining Hubble's parameter and the masses of galaxies from the gravitational lens effect}",
  journal = {Monthly Notices of the Royal Astronomical Society},
  year = 1964,
  volume = {128},
  pages = {307},
  doi = {10.1093/mnras/128.4.307},
  adsurl = {https://ui.adsabs.harvard.edu/abs/1964MNRAS.128..307R}
}

@ARTICLE{treu,
  author = {{Treu}, T. and {Marshall}, P.~J.},
  title = "{Time delay cosmography}",
  journal = {Astronomy and Astrophysics Review},
  year = 2016,
  volume = {24},
  pages = {11},
  doi = {10.1007/s00159-016-0096-8},
  adsurl = {https://ui.adsabs.harvard.edu/abs/2016A&ARv..24...11T}
}

@ARTICLE{bode,
  author = {{Bode}, P. and {Ostriker}, J. P. and {Turok}, N.},
  title = "{Halo Formation in Warm Dark Matter Models}",
  journal = {Astrophysical Journal},
  year = 2001,
  volume = {556},
  pages = {93-107},
  doi = {10.1086/321541},
  adsurl = {https://ui.adsabs.harvard.edu/abs/2001ApJ...556...93B}
}

@ARTICLE{viel,
  author = {{Viel}, M. and {Becker}, G. D. and {Bolton}, J. S. and {Haehnelt}, M. G.},
  title = "{Warm dark matter as a solution to the small scale crisis: New constraints from high redshift Lyman-\alpha forest data}",
  journal = {Physical Review D},
  year = 2013,
  volume = {88},
  pages = {043502},
  doi = {10.1103/PhysRevD.88.043502},
  adsurl = {https://ui.adsabs.harvard.edu/abs/2013PhRvD..88d3502V}
}

@ARTICLE{dayal,
  author = {{Dayal}, P. and {Lovisari}, M. and {Borgani}, S. and {Choudhury}, T. R.},
  title = "{Warm dark matter constraints from the JWST}",
  journal = {Monthly Notices of the Royal Astronomical Society},
  year = 2024,
  volume = {528},
  pages = {2784-2797},
  doi = {10.1093/mnras/stae071},
  adsurl = {https://ui.adsabs.harvard.edu/abs/2024MNRAS.528.2784D}
}

@article{ mohit,
	author = {{Yadav, Mohit} and {Sarkar, Tapomoy Guha}},
	title = {Probing decaying dark matter using the post-reionization HI 21-cm signal},
	DOI= "10.1140/epjc/s10052-025-15055-3",
	url= "https://doi.org/10.1140/epjc/s10052-025-15055-3",
	journal = {Eur. Phys. J. C},
	year = 2025,
	volume = 85,
	number = 11,
	pages = "1337",
}

@article{Riess2016,
  author = {Riess, A.~G. and others},
  title = {A 2.4\% Determination of the Local Value of the Hubble Constant},
  journal = {Astrophys. J.},
  volume = {826},
  pages = {56},
  year = {2016},
  doi = {10.3847/0004-637X/826/1/56}
}

@article{DiValentino2021,
  author = {Di Valentino, Eleonora and others},
  title = {In the Realm of the Hubble tension — A Review of Solutions},
  journal = {Class. Quantum Grav.},
  volume = {38},
  pages = {153001},
  year = {2021},
  doi = {10.1088/1361-6382/ac086d}
}

@article{Verde2019,
  author = {Verde, Licia and Treu, Tommaso and Riess, Adam G.},
  title = {Tensions between the Early and Late Universe},
  journal = {Nature Astronomy},
  volume = {3},
  pages = {891},
  year = {2019},
  doi = {10.1038/s41550-019-0902-0}
}

@article{Hildebrandt2017,
  author = {Hildebrandt, Hendrik and others},
  title = {KiDS-450: Cosmological parameter constraints from tomographic weak gravitational lensing},
  journal = {Mon. Not. Roy. Astron. Soc.},
  volume = {465},
  pages = {1454},
  year = {2017},
  doi = {10.1093/mnras/stw2805}
}

@article{Asgari2021,
  author = {Asgari, Marika and others},
  title = {KiDS-1000 Cosmology: Cosmic shear constraints and comparison between two point statistics},
  journal = {Astron. Astrophys.},
  volume = {645},
  pages = {A104},
  year = {2021},
  doi = {10.1051/0004-6361/202039070}
}

@article{Abdalla2022,
  author = {Abdalla, Elcio and others},
  title = {Cosmology Intertwined II: The Hubble Constant Tension},
  journal = {Astropart. Phys.},
  volume = {131},
  pages = {102605},
  year = {2022}
}

@article{Heymans2021,
  author = {Heymans, Catherine and others},
  title = {KiDS-1000 Cosmology: Multi-probe weak gravitational lensing},
  journal = {Astron. Astrophys.},
  volume = {646},
  pages = {A140},
  year = {2021}
}

@ARTICLE{2005PhR...405..279B,
       author = {{Bertone}, Gianfranco and {Hooper}, Dan and {Silk}, Joseph},
        title = "{Particle dark matter: Evidence, candidates and constraints}",
      journal = {Physics Reports},
         year = 2005,
        month = jan,
       volume = {405},
       number = {5-6},
        pages = {279-390},
          doi = {10.1016/j.physrep.2004.08.031},
archivePrefix = {arXiv},
       eprint = {hep-ph/0404175},
 primaryClass = {hep-ph},
       adsurl = {https://ui.adsabs.harvard.edu/abs/2005PhR...405..279B}
}

@ARTICLE{2015PNAS..11212249W,
       author = {{Weinberg}, David H. and {Bullock}, James S. and {Governato}, Fabio and {Kuzio de Naray}, Rachel and {Peter}, Annika H.~G.},
        title = "{Cold dark matter: controversies on small scales}",
      journal = {Proceedings of the National Academy of Science},
         year = 2015,
        month = oct,
       volume = {112},
       number = {40},
        pages = {12249-12255},
          doi = {10.1073/pnas.1308716112},
archivePrefix = {arXiv},
       eprint = {1306.0913},
 primaryClass = {astro-ph.CO},
       adsurl = {https://ui.adsabs.harvard.edu/abs/2015PNAS..11212249W}
}

@ARTICLE{2017ARA&A..55..343B,
       author = {{Bullock}, James S. and {Boylan-Kolchin}, Michael},
        title = "{Small-Scale Challenges to the ΛCDM Paradigm}",
      journal = {Annual Review of Astronomy and Astrophysics},
         year = 2017,
        month = aug,
       volume = {55},
        pages = {343-387},
          doi = {10.1146/annurev-astro-091916-055313},
archivePrefix = {arXiv},
       eprint = {1707.04256},
 primaryClass = {astro-ph.CO},
       adsurl = {https://ui.adsabs.harvard.edu/abs/2017ARA&A..55..343B}
}

@ARTICLE{2000PhRvL..84.3760S,
       author = {{Spergel}, David N. and {Steinhardt}, Paul J.},
        title = "{Observational Evidence for Self-Interacting Cold Dark Matter}",
      journal = {Physical Review Letters},
         year = 2000,
        month = apr,
       volume = {84},
       number = {17},
        pages = {3760-3763},
          doi = {10.1103/PhysRevLett.84.3760},
archivePrefix = {arXiv},
       eprint = {astro-ph/9909386},
 primaryClass = {astro-ph},
       adsurl = {https://ui.adsabs.harvard.edu/abs/2000PhRvL..84.3760S}
}

@ARTICLE{2000PhRvL..85.1158H,
       author = {{Hu}, Wayne and {Barkana}, Rennan and {Gruzinov}, Andrei},
        title = "{Fuzzy Cold Dark Matter: The Wave Properties of Ultralight Particles}",
      journal = {Physical Review Letters},
         year = 2000,
        month = aug,
       volume = {85},
       number = {6},
        pages = {1158-1161},
          doi = {10.1103/PhysRevLett.85.1158},
archivePrefix = {arXiv},
       eprint = {astro-ph/0003365},
 primaryClass = {astro-ph},
       adsurl = {https://ui.adsabs.harvard.edu/abs/2000PhRvL..85.1158H}
}

@ARTICLE{2001ApJ...546L..77C,
       author = {{Cen}, Renyue},
        title = "{Decaying Cold Dark Matter Model and Small-Scale Power}",
      journal = {Astrophysical Journal Letters},
         year = 2001,
        month = jan,
       volume = {546},
        pages = {L77-L80},
          doi = {10.1086/318847},
archivePrefix = {arXiv},
       eprint = {astro-ph/0005206},
 primaryClass = {astro-ph},
       adsurl = {https://ui.adsabs.harvard.edu/abs/2001ApJ...546L..77C}
}

@ARTICLE{2008PhRvD..77f3514B,
       author = {{Borzumati}, Francesca and {Bringmann}, Torsten and {Ullio}, Piero},
        title = "{Dark matter from late decays and the small-scale structure problems}",
      journal = {Physical Review D},
         year = 2008,
       volume = {77},
          eid = {063514},
        pages = {063514},
          doi = {10.1103/PhysRevD.77.063514},
archivePrefix = {arXiv},
       eprint = {hep-ph/0701007},
 primaryClass = {hep-ph},
       adsurl = {https://ui.adsabs.harvard.edu/abs/2008PhRvD..77f3514B}
}

@ARTICLE{2010PhRvD..81j3501P,
       author = {{Peter}, Annika H.~G. and {Benson}, Andrew J.},
        title = "{Dark-matter decays and self-gravitating halos}",
      journal = {Physical Review D},
         year = 2010,
       volume = {81},
          eid = {103501},
        pages = {103501},
          doi = {10.1103/PhysRevD.81.103501},
archivePrefix = {arXiv},
       eprint = {1003.0419},
 primaryClass = {astro-ph.CO},
       adsurl = {https://ui.adsabs.harvard.edu/abs/2010PhRvD..81j3501P}
}

@article{gp1,
  title = {Nonparametric reconstruction of the dark energy equation of state from diverse data sets},
  author = {Holsclaw, Tracy and Alam, Ujjaini and Sans\'o, Bruno and Lee, Herbie and Heitmann, Katrin and Habib, Salman and Higdon, David},
  journal = {Phys. Rev. D},
  volume = {84},
  issue = {8},
  pages = {083501},
  numpages = {18},
  year = {2011},
  month = {Oct},
  publisher = {American Physical Society},
  doi = {10.1103/PhysRevD.84.083501},
  url = {https://link.aps.org/doi/10.1103/PhysRevD.84.083501}
}

@ARTICLE{gp2,
       author = {{Seikel}, Marina and {Clarkson}, Chris and {Smith}, Mathew},
        title = "{Reconstruction of dark energy and expansion dynamics using Gaussian processes}",
      journal = {\jcap},
         year = 2012,
        month = jun,
       volume = {2012},
       number = {6},
          eid = {036},
        pages = {036},
          doi = {10.1088/1475-7516/2012/06/036},
archivePrefix = {arXiv},
       eprint = {1204.2832},
 primaryClass = {astro-ph.CO},
       adsurl = {https://ui.adsabs.harvard.edu/abs/2012JCAP...06..036S}
}

@ARTICLE{gp3,
       author = {{Shafieloo}, Arman and {Kim}, Alex G. and {Linder}, Eric V.},
        title = "{Gaussian process cosmography}",
      journal = {\prd},
         year = 2012,
        month = jun,
       volume = {85},
       number = {12},
          eid = {123530},
        pages = {123530},
          doi = {10.1103/PhysRevD.85.123530},
archivePrefix = {arXiv},
       eprint = {1204.2272},
 primaryClass = {astro-ph.CO},
       adsurl = {https://ui.adsabs.harvard.edu/abs/2012PhRvD..85l3530S}
}

@ARTICLE{gp4,
       author = {{Mukherjee}, Purba and {Sen}, Anjan Ananda},
        title = "{Model-independent cosmological inference post DESI DR1 BAO measurements}",
      journal = {\prd},
         year = 2024,
        month = dec,
       volume = {110},
       number = {12},
          eid = {123502},
        pages = {123502},
          doi = {10.1103/PhysRevD.110.123502},
archivePrefix = {arXiv},
       eprint = {2405.19178},
 primaryClass = {astro-ph.CO},
       adsurl = {https://ui.adsabs.harvard.edu/abs/2024PhRvD.110l3502M}
}

@book{gp5,
author = {Rasmussen, Carl and Williams, Christopher},
year = {2005},
month = {11},
pages = {},
title = {Gaussian Processes for Machine Learning},
publisher = {MIT Press},
isbn = {9780262256834},
doi = {10.7551/mitpress/3206.001.0001}
}

\end{document}